\documentclass[conference]{IEEEtran}

\usepackage{tikz}
\usepackage{amsmath}
\usepackage{amssymb}
\usepackage[all]{nowidow}
\usepackage{siunitx}
\usepackage[linesnumbered,ruled,vlined]{algorithm2e}
\usepackage{multirow}
\usepackage{amsmath}

\usepackage{enumitem}
\usepackage{tcolorbox}
\usepackage{caption}
\usepackage{subcaption}
\usepackage{colortbl}
\usepackage{diagbox, eqparbox, hhline}
\usepackage{hyperref}

\usepackage{booktabs}
\usepackage{rotating}
\usepackage{xurl}
\usepackage{tabularx}
\newcolumntype{Y}{>{\raggedright\arraybackslash}X}
\usepackage{collcell}

\usepackage{graphicx}
\newcommand*\rot{\rotatebox{90}}
\usepackage{pifont} 
\newcommand{\xmark}{\ding{55}}%
\usepackage{tikz}
\newcommand{\emptypie}{%
\begin{tikzpicture}
 \draw (0,0) circle (.75ex);
\end{tikzpicture}%
}
\newcommand{\pie}[1]{%
\begin{tikzpicture}
 \draw (0,0) circle (.75ex);\fill[rotate=90] (.75ex,0) arc (0:#1:.75ex) -- (0,0) -- cycle;
\end{tikzpicture}%
}

\usepackage{xcolor}
\newcommand*\circled[1]{\tikz[baseline=(char.base),scale=0.8,transform shape]{
            \node[shape=circle,fill,inner sep=.8pt] (char) {\footnotesize\textcolor{white}{#1}};}}

\newcommand{\refappendix}[1]{\hyperref[#1]{Appendix~\ref*{#1}}}

\definecolor{Gray}{gray}{0.9}

\newif{\ifanonymous}
\anonymousfalse{} 

\newlist{rqlist}{enumerate}{3}
\setlist[rqlist]{label=\textbf{RQ\arabic*}.,leftmargin=30pt,ref=RQ\arabic*}


\usepackage{framed}
\renewenvironment{leftbar}[2][\hsize]
{
    \def\FrameCommand
    {
        {\color{#2}\vrule width 3pt}
        \hspace{0pt}
    }
    \MakeFramed{\hsize#1\advance\hsize-\width\FrameRestore}
}
{\endMakeFramed}

\newcounter{finding}
\newcommand{\finding}[1]{\vspace{4pt}
\refstepcounter{finding}
\begin{leftbar}{gray}
\noindent\textbf{Finding \hyperref[tab:findings-mapping]{F\arabic{finding}}} - \textit{#1}\quad
\end{leftbar}}
\makeatletter
\def\findingautorefname{F\@gobble}
\makeatother

\newcounter{recommendation}
\newcommand{\recommendation}[1]{\vspace{4pt}
\refstepcounter{recommendation}
\begin{leftbar}{black}
\noindent\textbf{Recommendation \hyperref[tab:findings-mapping]{R\arabic{recommendation}}} - \textit{#1}\quad
\end{leftbar}}
\makeatletter
\def\recommendationautorefname{R\@gobble}
\makeatother

\newcommand{\NumTopWebsites}{100k\xspace}

\begin{document}
\title{Longitudinal Adoption and Deprecation of the Privacy Sandbox Web APIs}

\ifanonymous
\author{\em Anonymous Authors}
\else
\author{\IEEEauthorblockN{Yohan Beugin}
\IEEEauthorblockA{University of Wisconsin-Madison\\
ybeugin@cs.wisc.edu}
\and
\IEEEauthorblockN{Paul Barford}
\IEEEauthorblockA{University of Wisconsin-Madison\\
pb@cs.wisc.edu}
\and
\IEEEauthorblockN{Patrick McDaniel}
\IEEEauthorblockA{University of Wisconsin-Madison\\
mcdaniel@cs.wisc.edu}}
\fi

\maketitle

\begin{abstract}
While several web actors have been trying to reduce web tracking for years, it remains unclear how to achieve both desirable levels of utility and privacy. In 2019, Google launched the Privacy Sandbox initiative to balance that trade-off and find privacy alternatives to common use cases such as advertising.
Yet, in late 2025, Google canceled the project and deprecated most of the newly introduced APIs. Despite its end, the Privacy Sandbox represents a unique opportunity to learn about how the ecosystem reacted to the proposed changes.
In this paper, we present a longitudinal measurement and analysis study of the Privacy Sandbox APIs to characterize their adoption and deprecation over the past seven years by different web actors.
Leveraging historical HTTP Archive crawls and public Chrome telemetry data, we offer the largest study of its kind into the prevalence of each Privacy Sandbox feature, during their entire respective lifetime (5+ years for some), on popular websites (CrUX top \NumTopWebsites), and as experienced by Chrome users during their browsing journey.
Our results showcase an adoption that remained limited and uneven across the years; only few web actors implemented very specific APIs, and in disparate manners.
We motivate our interpretation of these results by considering the incentives (interest, resources, timeline, etc.) and risks (potential trade-offs, privacy violations, and legal exposure, etc.) for these actors. Finally, our analysis also yields a few actionable recommendations for the next generation of web privacy-enhancing technologies.
\end{abstract}


%
\IEEEpeerreviewmaketitle

\section{Introduction}
Despite web browsers deploying several privacy solutions over the years, web tracking continues to be a pervasive practice today, as it remains unclear how to achieve both desirable levels of utility and privacy for web actors and users~\cite{vekariaSoKAdvancesOpen2025}.
In August 2019, Google launched the Privacy Sandbox initiative to reduce \emph{``cross-site and cross-app tracking while helping to keep online content and services free for all''}~\cite{googleWhatPrivacySandbox2021}.
As part of this project, Google proposed several alternative mechanisms to restrict the use of third-party cookies (TPC) and fingerprinting techniques on Chrome while still supporting advertising, conversion measurement, and fraud detection use cases~\cite{googlePotentialUsesPrivacy2019}.

Google's initiative represents one of the largest experiments on deploying web privacy-enhancing techniques in recent years and on exploring with the community mechanisms to balance such privacy and utility trade-offs.
However, in October 2025, Google ultimately announced the continuing support for TPC in Chrome and the deprecation of most of the newly introduced APIs~\cite{chavezUpdatePlansPrivacy2025}.
Despite its end, the Privacy Sandbox experiment offers a unique perspective into how the ecosystem used (and abandoned) the proposed changes.

In this paper, we present a longitudinal measurement and analysis study of the Privacy Sandbox to shed light on the adoption and deprecation over the years by different web actors of its technologies when they were available in Chrome.
Through such observations, we obtain valuable insights about how this global initiative was adopted or not, which also yield actionable recommendations for the next generation of web privacy-enhancing technologies.
Our study leverages historical and public datasets to offer the largest study of its kind into the prevalence of each Privacy Sandbox feature over its lifetime in Chrome (5+ years for some).
Among other sources, we query historical crawls from the HTTP Archive to analyze the adoption and/or deprecation of each proposal on the top \NumTopWebsites websites (per CrUX rank).
We also use public and aggregated Chrome telemetry to gain a view of the Privacy Sandbox features encountered by Chrome users while browsing.

\begin{table*}[!ht]
  \centering
\caption{Study findings and recommendations (more details in each respective section).}
\label{tab:findings-mapping}
\setlength\tabcolsep{3pt}
\begin{tabular*}{\textwidth}{cllcl}
\toprule
\multicolumn{2}{c}{\textbf{Finding}} & \multicolumn{1}{c}{\textbf{Metric}} & \multicolumn{2}{c}{\textbf{Recommendation}}\\
\midrule

\multicolumn{3}{l}{\underline{\textit{API Adoption}} (\autoref{sec:adoption})}\\
(\autoref{finding:adoption-uneven}) & \begin{tabular}[l]{@{}l@{}}API adoption is highly uneven.\end{tabular} & \begin{tabular}[l]{@{}l@{}}5 only $>11.7\%$ page loads.\end{tabular} & (\autoref{rec:api-use-cases-value}) & \begin{tabular}[l]{@{}l@{}}
Future efforts should concentrate on advertising use cases. \end{tabular}\\

(\autoref{finding:adoption-rise-plateau}) & \begin{tabular}[l]{@{}l@{}} Adoption rose after introduction of APIs,\\ but peaked and stagnated by Q3 2024. \end{tabular}  & \begin{tabular}[l]{@{}l@{}}See Figures~\ref{fig:telemetry},~\ref{fig:telemetry-log},~\ref{fig:ha-blink}, and~\ref{fig:ha-blink-log}. \end{tabular} & (\autoref{rec:advertising-apis}) & \begin{tabular}[l]{@{}l@{}}
The community will be reluctant to adopt future privacy\\proposals from Google as well as other actors. \end{tabular}\\

\multicolumn{3}{l}{\underline{\textit{API Adopters}} (\autoref{sec:adopters})}\\
(\autoref{finding:adopters-attestation}) & \begin{tabular}[l]{@{}l@{}}The majority of adopters were active\\ starting in the early days of the initiative.\end{tabular} & \begin{tabular}[l]{@{}l@{}}295 attestations, 73 \texttt{RWS} sets, top\\ adopters consistent years apart.
\end{tabular} & (\autoref{rec:early-testers})  & \begin{tabular}[l]{@{}l@{}}
Securing long term commitment from early adopters is\\
important to encourage further adoption by others.
\end{tabular} \\

\multicolumn{3}{l}{\underline{\textit{Supported APIs}} (\autoref{sec:left})}\\

(\autoref{finding:remaining-apis}) & \begin{tabular}[l]{@{}l@{}} Most supported APIs have low usage.\end{tabular} & $\leq5.2\%$ of page loads.& 
\multirow{2}*{(\autoref{rec:external-incentives})}  &\multirow{2}*{\begin{tabular}[l]{@{}l@{}}
External incentives and economics changes may be needed\\
to encourage adoption of the supported APIs and future ones.
\end{tabular}}\\

(\autoref{finding:remaining-chips}) & \begin{tabular}[l]{@{}l@{}} \texttt{CHIPS} adoption is driven by actors\\ with large coverage of the web. \end{tabular} & \begin{tabular}[l]{@{}l@{}}$\sim34.5\%$ of page loads, but\\ only $6.7\%$ of TPC on average.\end{tabular} &  & \\

(\autoref{finding:ua-ch-abuse}) & \begin{tabular}[l]{@{}l@{}} \texttt{UA Client Hints} is being\\ abused for browser fingerprinting.\end{tabular} & $>62\%$ of page loads. & (\autoref{rec:apis-value}) & \begin{tabular}[l]{@{}l@{}}Continuing use of \texttt{UA Client Hints} should be justified\\ and user-friendly controls to disable the API must be provided.\end{tabular}\\

\midrule
\multicolumn{5}{c}{\textbf{Other Recommendations}}\\
\midrule
\multicolumn{2}{l}{\underline{\textit{Background}} (\autoref{sec:background})} & & \multicolumn{2}{l}{\underline{\textit{Discussion}} (\autoref{sec:discussion})}\\

(\autoref{rec:eval-before-deploy})& \multicolumn{2}{l}{Proposals should be precisely evaluated before deployment.}&(\autoref{rec:consent})& Consent and legal implications should be part of API design.\\

(\autoref{rec:attestation})& \multicolumn{2}{l}{Proposals should not be exploitable from a technical aspect.}& (\autoref{rec:privacy-by-default})& Privacy protections should be deployed by default to users.\\

\bottomrule
\end{tabular*}
\end{table*}

Our results (overview in~\autoref{tab:findings-mapping}) showcase a general adoption of the Privacy Sandbox that remained limited and very uneven across the years depending on the considered API. We observe that most adoption rates reached about their maximum and stagnated before Google postponed the deprecation of TPC. We find that the majority of adopters have been present since the early days of the initiative and that significant efforts appear to have been deployed by them to test and implement the proposed mechanisms. However, we detect continuing use of the \texttt{UA Client Hints} API which weakens the provided protections of UA reduction efforts, usage of \texttt{UA Client Hints} requires further justification due to its browser fingerprinting risks.
Finally, without TPC deprecation in Chrome, we expect that other external incentives will be needed to encourage further adoption of the rest of the supported APIs.

Our analysis sheds light on the overall limited and uneven adoption of the Privacy Sandbox APIs by a few specific web actors.
More broadly, the Privacy Sandbox illustrates the limitations of \textit{``fix it in the browser''} remedies and disparities across browser vendors. Despite Google's initial objective of \textit{``making the web more private''}, today, tracking and TPC limitations in Chrome remain largely opt-in, while they have been enabled by default on other browsers like Brave, Firefox, and Safari.
Our contributions are as following:
\begin{itemize}
  \item We perform the largest longitudinal measurement analysis study of its kind on the adoption (HTTP Archive) and usage (Chrome Telemetry) of the Privacy Sandbox.

  \item We report on which APIs were adopted, by whom, and how, and interpret our findings within their broader context, i.e., with respect to the development timeline of the proposals, their prior analyses, but also the incentives and risk factors that web actors must consider, and the protections deployed by other browsers in the meantime.
  
  \item We highlight privacy issues and violations, such as API call prior to user consent, continuing use of \texttt{UA Client Hints}, and its abuse for browser fingerprinting.
  
  \item We offer a few actionable recommendations for the next generation of web privacy-enhancing technologies.

  \item Finally, we open-source our entire measurement approach as a reproducible artifact to encourage further exploration and extensions of this work (see~\refappendix{sec:open-science}).
\end{itemize}

\section{Background on the Privacy Sandbox}
\label{sec:background}

Next, we provide details about the Privacy Sandbox and its proposals (with related adoption work described in~\autoref{sec:related-work}). For more background on web tracking overall, we refer to the following SoK from Vekaria et al.~\cite{vekariaSoKAdvancesOpen2025}.

\begin{figure}[!ht]
\begin{tcolorbox}
\small
\subsection*{Timeline Overview}

\noindent
\textbf{Aug 2019.} Launch of the Privacy Sandbox~\cite{schuhBuildingMorePrivate2019a,googlePotentialUsesPrivacy2019}.

\noindent
\textbf{Jan 2020.} Google plans phase-out support for TPC within two years and first origin trials (OT) by end 2020~\cite{schuhBuildingMorePrivate2020}.

\noindent
\textbf{Jan 2021.} UK's Competition and Markets Authority (CMA) investigates Google's proposals in Chrome~\cite{competitionandmarketsauthorityInvestigationGooglesPrivacy2025}.

\noindent
\circled{1} \textbf{Apr 2022.} Start of unified OT in Chrome Beta~\cite{googleRelevanceMeasurementUnified2025}.

\noindent
\circled{2} \textbf{Jul 2022.} TPC deprecation delayed to Q3 or Q4 2024, expansion of the OT to Chrome stable users in August, and general availability for APIs in Q3 2023~\cite{chavezExpandingTestingPrivacy2022}.

\noindent
\circled{3} \textbf{Oct 2022.} 5\% of stable users are now in the OT~\cite{googleIncreasingPrivacySandbox2022}.

\noindent
\textbf{Apr 2023.} Enrollment required for \texttt{Attribution Reporting}, \texttt{Fenced Frames}, \texttt{Private Aggregation}, \texttt{Protected Audience}, \texttt{Shared Storage}, \texttt{Topics} and attest no cross-site tracking~\cite{googleDeveloperEnrollmentPrivacy2023}.

\noindent
\circled{4} \textbf{May 2023.} General APIs availability starting July 2023, TPC deprecation for 1\% of users starting Q1 2024. Full deprecation still planned for Q3 or Q4 2024~\cite{chavezNextStagesPrivacy2023}.

\noindent
\circled{5} \textbf{Sep 2023.} \texttt{Attribution Reporting}, \texttt{Fenced Frames}, \texttt{Private Aggregation}, \texttt{Protected Audience}, \texttt{Shared Storage}, and \texttt{Topics} ship by default in Chrome 115~\cite{chavezPrivacySandboxWeb2023}.

\noindent
\circled{6} \textbf{Jan 2024.} TPC restricted for 1\% of Chrome users~\cite{duttonThirdpartyCookiesRestricted2024}.

\noindent
\circled{7} \textbf{Apr 2024.} Google announces that, after all, TPC will not be deprecated in Chrome by the end of Q4 2024~\cite{googleUpdatePlanPhaseout2024}.

\noindent
\circled{8} \textbf{Jul 2024.} No TPC deprecation considered anymore, new TPC consent mechanism planned instead~\cite{chavezNewPathPrivacy2024}.

\noindent
\textbf{Apr 2025.} Google restates that TPC will not be deprecated and announces not rolling out the new TPC standalone prompt. Unclear future for Privacy Sandbox APIs~\cite{chavezNextStepsPrivacy2025}.

\noindent
\circled{9} \textbf{Oct 2025.} Most APIs are marked for deprecation and TPC will continue to be supported in Chrome. CMA releases Google from its commitments~\cite{chavezUpdatePlansPrivacy2025}.

\noindent
\circled{10} \textbf{Jan 2026.} Deprecation of \texttt{Attribution Reporting}, \texttt{Private Aggregation}, \texttt{Protected Audience}, \texttt{RWS}, \texttt{Shared Storage}, and \texttt{Topics} in Chrome 144~\cite{googleIntentDeprecateRemove2025,googleDeprecateRemoveShared2025,googleDeprecateRemoveRelated2025,googleDeprecateRemoveProtected2025,googleDeprecateRemovePrivate2025,googleDeprecateRemoveAttribution2025} and complete removal in Chrome 152 in August 2026.
\end{tcolorbox}

\caption{Privacy Sandbox timeline overview}
\label{aside:timeline}
\end{figure}

\begin{table*}[!ht]
\centering
\caption{Overview of the 13 main Privacy Sandbox proposals (in chronological order) measured in this paper. Note: APIs without a deprecated date are still supported in Chrome. See \autoref{sec:related-work} for an overview of related work on adoption.}
\label{tab:apis}
\setlength\tabcolsep{1pt}
\newcolumntype{K}[1]{>{\centering\arraybackslash}m{#1}}
\begin{tabular}{@{}K{4.2cm}K{3.5cm}K{1cm}K{1cm}K{1.4cm}K{2.5cm}K{3.7cm}@{}}
\toprule
\multirow{2}{*}{\textbf{API [Explainer]}} &  \multirow{2}{*}{\textbf{Use Case}} &  \multicolumn{3}{c}{\textbf{Dates}} &  \multicolumn{2}{c}{\textbf{Related Work}} \\
\cmidrule(lr){3-5} \cmidrule(l){6-7}  &  & Testing & Shipping & Deprecated  & Adoption & Other \\
\midrule
\texttt{FLoC} \cite{WICGFloc2019} & Ad targeting & - & 2021-03 & 2021-08 &  \cite{vekariaWebAlmanacPrivacy2024} &  \cite{ravichandranEvaluationCohortAlgorithms2020,epastoClusteringPrivateInterestbased2021,rescorlaTechnicalCommentsFLoC2021,berkePrivacyLimitationsInterestbased2022,kessibiComplementaryUtilityPrivacy2022,turatiLocalitySensitiveHashingDoes2023}\\

\texttt{UA Client Hints} \cite{WICGUaclienthints2018} & Access reduced UA features & - & 2021-03 & - &  \cite{senolUnveilingImpactUserAgent2023,wieflingPrivacyMeasureTurned2024,vekariaWebAlmanacPrivacy2024,ilgazInvestigatingHighEntropyClient2025} &  \cite{intumwayaseUARadarExploringImpact2023}\\

\texttt{FedCM} \cite{W3cfedidFedCM2020} & Single Sign-On login & 2022-06 & 2022-11 & - & \cite{gribRiseFallGoogle2026} &  \cite{westersSingleSignOnPrivacy2024}\\

\texttt{CHIPS} \cite{PrivacycgCHIPS2021} & Opt-in partitioned cookies & 2022-03 & 2023-05 & - &  \cite{beuginWebAlmanacCookies2024,beuginWebAlmanacCookies2026,zollnerFirstLookCookies2025a,gribRiseFallGoogle2026} & -\\

\texttt{Private State Tokens} \cite{WICGTrusttokenapi2019} & Ad fraud prevention & 2020-07 & 2023-07 & - & \cite{gribRiseFallGoogle2026} &  \cite{aliNavigatingMurkyWaters2023}\\

\texttt{RWS} \cite{WICGFirstpartysets2019} & Cross-site relationships & 2021-03 & 2023-07 & 2026-01 &  \cite{mcquistinFirstLookRelated2024,beuginWebAlmanacCookies2024,johnsonUnearthingPrivacyEnhancingAd2024} & -\\

\texttt{Attribution Reporting} \cite{WICGAttributionreportingapi2019} & Ad conversion measurement & 2022-04 & 2023-07 & 2026-01 &  \cite{vekariaWebAlmanacPrivacy2024,gribRiseFallGoogle2026} &  \cite{langPrivacySandboxAggregation2024,aksuSummaryReportsOptimization2024,delaneyDifferentiallyPrivateAd2024,xiaoClickCompromiseOnline2025}\\

\texttt{Private Aggregation} \cite{PatcgindividualdraftsPrivateaggregationapi2022} & Cross-site measurement & 2022-04 & 2023-07 & 2026-01 & \cite{gribRiseFallGoogle2026} & -\\

\texttt{Protected Audience} \cite{WICGTurtledove2020} & Ad retargeting & 2022-04 & 2023-07 & 2026-01 &  \cite{ruminskiFindingsEarlyFledge2022,philipsePostThirdPartyCookies2024,calderonioFledgingWillContinue2024,vekariaWebAlmanacPrivacy2024,johnsonUnearthingPrivacyEnhancingAd2024,johnsonAdventPrivacycentricDigital2024,gribRiseFallGoogle2026} &  \cite{aliNavigatingMurkyWaters2023,thomsonProtectedAudiencePrivacy2024,longEvaluatingGooglesProtected2024}\\

\texttt{Topics} \cite{PatcgindividualdraftsTopics2022a} & Ad targeting & 2022-04 & 2023-07 & 2026-01 &  \cite{vekariaWebAlmanacPrivacy2024,johnsonUnearthingPrivacyEnhancingAd2024,johnsonAdventPrivacycentricDigital2024,philipsePostThirdPartyCookies2024,vernaUnderstandingTopicsAPI2025,gribRiseFallGoogle2026} &  \cite{epastoMeasuresCrosssiteReidentification2022,careyMeasuringReidentificationRisk2023,thomsonPrivacyAnalysisGoogles2023,jhaReIdentificationAttacksTopics2024,alvimPrivacyUtilityTradeoffTopics2024,beuginPublicReproducibleAssessmentTopics2024,beuginInterestDisclosingMechanismsAdvertising2024,dickDifferentiallyPrivateSynthetic2025}\\

\texttt{Fenced Frames} \cite{WICGFencedframe2020} & Isolated iframes & 2022-05 & 2023-07 & - & \cite{gribRiseFallGoogle2026} & -\\

\texttt{Storage Access} \cite{StorageAccessAPI2020} & Legitimate third-party storage & - & 2023-07 & - & \cite{gribRiseFallGoogle2026} & -\\

\texttt{Shared Storage} \cite{WICGSharedstorage2021} & Cross-site restricted storage & 2022-08 & 2023-07 & 2026-01 & \cite{gribRiseFallGoogle2026} &  \cite{nisenoffExploitingSharedStorage2025}\\

\bottomrule
\end{tabular}
\end{table*}
\subsection{Objectives \& Timeline}

Google launched the Privacy Sandbox initiative in August 2019 to reduce the use of identifiers while still supporting cross-site and cross-app use cases such as ad targeting, conversion measurement, and fraud detection~\cite{googlePotentialUsesPrivacy2019,schuhBuildingMorePrivate2019a}.
The project was initially created for the web with more than 25 features developed to deprecate TPC in Chrome by 2022. Proposals were later extended to the Android platform (6 features total) to discontinue the Advertising ID and limit the access scope of third-party library and SDKs~\cite{googlePrivacySandboxFeature2025}.

Over the years, some of these new web mechanisms were directly implemented and shipped in Chrome. Google often did so without reaching consensus with other browsers (e.g., Brave, Mozilla, Safari) and despite external evaluations uncovering limitations in the privacy claims being made~\cite{beuginTechnicalReportSandstorm2025}.
As evidenced by the timeline overview in~\autoref{aside:timeline}, the initial deprecation date for TPC support in Chrome was postponed several times. Ultimately, Google ended the project in October 2025 by announcing the deprecation of all but 6 of the APIs and the continuing support for TPC in Chrome~\cite{chavezUpdatePlansPrivacy2025,googlePrivacySandboxFeature2025}.

\subsection{Privacy Analyses of Proposals}
\label{sec:related-work-privacy}
Given its potential impact on the online ecosystem and users' privacy, the Privacy Sandbox has attracted the attention of many researchers across the years.
Google itself released several whitepapers and preliminary analyses of some proposals~\cite{ravichandranEvaluationCohortAlgorithms2020,epastoClusteringPrivateInterestbased2021,epastoMeasuresCrosssiteReidentification2022,careyMeasuringReidentificationRisk2023,langPrivacySandboxAggregation2024,aksuSummaryReportsOptimization2024,delaneyDifferentiallyPrivateAd2024}. However, these were often performed on private browsing history data or on public but out-of-domain datasets, with results reported in aggregate rather than showing the entire distribution, as well as at times with unclear methodology steps, assumptions, and limitations.
As a result, these issues led to active points of contention with the research community around the reproducibility of Google's claims and results. This is perhaps best illustrated by the evaluations of \texttt{Topics} for which Google released a synthetic dataset of users profiles only 2 years after it had already been shipped by default to Chrome users~\cite{dickDifferentiallyPrivateSynthetic2025}.
In response to Google attempting to standardize through the W3C some proposals, other browser vendors often published their own rebuttals and counter analyses~\cite{rescorlaTechnicalCommentsFLoC2021,thomsonPrivacyAnalysisGoogles2023,thomsonProtectedAudiencePrivacy2024}.
Finally, different academic researchers also produced a series of publications outlining the potential privacy leakage, cross-site tracking risk, and privacy-utility tension associated with these new mechanisms~\cite{berkePrivacyLimitationsInterestbased2022,kessibiComplementaryUtilityPrivacy2022,turatiLocalitySensitiveHashingDoes2023,jhaReIdentificationAttacksTopics2024,alvimPrivacyUtilityTradeoffTopics2024,beuginPublicReproducibleAssessmentTopics2024,beuginInterestDisclosingMechanismsAdvertising2024,longEvaluatingGooglesProtected2024,aliNavigatingMurkyWaters2023,intumwayaseUARadarExploringImpact2023,westersSingleSignOnPrivacy2024,nisenoffExploitingSharedStorage2025,xiaoClickCompromiseOnline2025}. We refer to~\autoref{tab:apis} for a more precise mapping between these works and each proposal.

\recommendation{Privacy guarantees of new web features should be precisely identified and evaluated in a reproducible manner before deployment to real users. Unbiased research can help to improve these mechanisms but need representative datasets and realistic testbeds.}\label{rec:eval-before-deploy}

\subsection{Web Proposals Overview}
\label{sec:background-web-apis}

More than 25 web features were developed by Google as part of the Privacy Sandbox. Among them, 18 made it to general availability (i.e., were shipped at some point by default in Chrome stable) and 7 others---that were never implemented---were abandoned following Google's deprecation announcement in October 2025. From the 18 features shipped in Chrome, 5 correspond to specific policy changes and were not, on their own, APIs that could be directly implemented by publishers and advertisers (e.g., bounce tracking mitigations, default partitioning, etc.).
The remaining 13 APIs are considered at the core of Google's efforts for the web; we focus on them in the rest of the paper. Next, we briefly describe their main use cases as envisioned by Google (past or present tense in these explanations mirror the deprecated or still supported status of each API, respectively). Please see~\autoref{tab:apis} for respective timelines and existing related work (described further in~\autoref{sec:related-work}) for each Privacy Sandbox API.
For more details on how each proposal actually works, see the following technical report by Beugin and McDaniel~\cite{beuginTechnicalReportSandstorm2025}.\\
\noindent\textbf{ \texttt{Attribution Reporting} (formerly \texttt{Conversion} \texttt{Measurement}).} This API allowed advertisers and ad tech providers to measure ad conversions without TPC. Chrome kept track of source (ad click, view, or touch) and trigger (purchase or conversion) events and linked them into coarse event-level or summary reports. Random noise and a time delay were added to reports to prevent linkage with real users' identities. Summary reports would be received in batches by the API caller that then must use an aggregation service to obtain the corresponding aggregated report.\\
\noindent\textbf{ \texttt{Cookies Having Independent Partitioned State (CHIPS).}} Web developers can opt in to save the cookies they set into partitioned storage, i.e., in a separate cookie jar per top-level site. Partitioned cookies allow for the legitimate use of TPC as they can only be accessed by the same third party and from the same top-level site where they were set in the first place. \texttt{CHIPS} is the only one of the 13 proposals also implemented in Firefox and Safari.\\
\noindent\textbf{\texttt{Federated Credential Management (FedCM).}} This mechanism enables identity providers to build an SSO-login infrastructure facilitated by compatible web browsers without the use of TPC or redirects. A relying party can allow users to sign in using the credentials of an account they have on a trusted third-party identity provider. The browser mediates the requests and information exchanged between both parties through the API.\\
\noindent\textbf{\texttt{Federated Learning of Cohorts (FLoC).}} One of the very first proposals from the project, \texttt{FLoC} aimed at enabling interest-based advertising without TPC. Chrome would assign users to a specific cohort every week based on their browsing history. A central server would enforce k-anonymity, checking that the cohort had enough users or merging it with other cohorts until satisfied. Finally, API callers would be able to observe the cohort IDs of users.\\
\noindent\textbf{\texttt{Fenced Frames.}} This API isolated \texttt{\small iframes}, with a strict boundary between the embedded content and the embedding page, preventing access to each other's context's \texttt{\small DOM}. This is achieved by the browser mapping the \texttt{Fenced Frame} to an opaque URL and restricting access to the resources available from inside it. They are intended to be used by \texttt{Protected Audience} or \texttt{Shared Storage} to prevent publishers from learning about an ad auction winner or the ad displayed.\\
\noindent\textbf{\texttt{Private Aggregation.}} Similar to \texttt{Attribution Reporting}, this proposal was meant to measure more general single and cross-site events, supporting other APIs, such as \texttt{Shared Storage} or \texttt{Protected Audience}.\\
\noindent\textbf{\texttt{Private State Tokens} (formerly \texttt{Trust Tokens}).} These tokens enable trust in a user’s authenticity to be conveyed in a cross-site manner to combat fraud and bots. An issuer website verifies that a user meets some definition of trustworthiness through a challenge (e.g., solving a CAPTCHA, logging into an account, performing a transaction, etc.). If completed successfully, the web browser is issued a token that proves cryptographically that the user is trusted by the issuer service. On the redeemer side, the website only needs to verify that the user has a valid and recent token from a trusted issuer and redeem these tokens as needed without challenging users again.\\
\noindent\textbf{\texttt{Protected Audience} (formerly \texttt{FLEDGE}).} This API provided a real-time bidding infrastructure for ad retargeting by relying on a new storage mechanism in the browser called \texttt{\small interest groups}. When users browsed the web and interacted with specific products, advertisers could request the users' browser to add them to a specific \texttt{\small interest group} that contains ads for the corresponding or similar product. Later, when a user would visit a website with an available ad space, Chrome would locally execute the ad auction logic for all \texttt{\small interest groups} users belong to and display the ad for the winning bid.\\
\noindent\textbf{\texttt{Related Website Sets (RWS)} (formerly \texttt{ First Party Sets}).} These represented declared relationships by companies among websites and services they owned. Chrome would vet and use these to decide when to allow or deny access to specific third-party storage.\\
\noindent\textbf{\texttt{Shared Storage.}} This API provided the needed data storage, processing, and sharing solution to allow cross-site information sharing, but without the use of identifiers. This was done by allowing unlimited, cross-site storage write access with privacy-preserving read access from only within a \texttt{\small worklet}. Worklets were to offer a secure environment for processing shared storage data, although direct data sharing with the associated context was not permitted.\\
\noindent\textbf{\texttt{Storage Access.}} Allows cross-origin content to request storage access as if they were in first-party context. Chrome automatically grants storage access in intra-\texttt{RWS} contexts (e.g., iframe whose embedded and top-level sites belong to the same \texttt{RWS}) or prompts users otherwise.\\
\noindent\textbf{\texttt{Topics.}} Successor to \texttt{FLoC}, this API required Chrome to classify websites visited by users into categories of interest picked from a taxonomy curated of sensitive topics by Google. Advertisers could then observe some of the recent top users’ topics and use them to perform an ad auction. Google claimed that users had plausible deniability due to noisy topics being sometimes returned and \textit{k}-anonymity protections from other users sharing their same interests.\\
\noindent\textbf{\texttt{User-Agent (UA) Client Hints.}} Several web browsers minimized identifying information shared through the \texttt{UA} header. This API allows servers to explicitly request access to features (also designated as hints) not exposed anymore by default by Chrome. The idea is that low-entropy \texttt{UA} hints are shared during the initial request, then servers can explicitly ask Chrome for more hints, technically allowing the browser to mediate access to higher entropy hints. However, the proposal fails to explain how to perform this mediation and Google decided to not pursue another project called \texttt{Privacy Budget}~\cite{westMikewestPrivacybudget2025} that may have been part of a solution. Thus, Chrome currently honors every request to obtain more \texttt{UA} hints.

\subsection{Attestation}
\label{sec:attestation-bkg}

In August 2023, Google started to mandate enrollment from all developers and third-parties that wanted to use and call the following APIs: \texttt{Attribution Reporting}, \texttt{Private Aggregation}, \texttt{Protected Audience}, \texttt{Shared Storage}, and \texttt{Topics}. This new requirement appeared to be Google's solution to address potential privacy violations and abuse of the APIs~\cite{patcg-individual-draftsFingerprintingThreatUsing2023}. It required API callers to host an attestation file at a specific \texttt{\small/.well-known/privacy-sandbox-attestation.json} URL under the domain the calls would originate from and to certify that they would not attempt re-identifying users~\cite{googlePrivacysandboxAttestation2023}.
On the other side, Chrome contained an allow-list to verify if an origin triggering one of these APIs was authorized to do so before returning the result.

\recommendation{Rather than using an attestation mechanism with unclear enforcement and legal outcomes if violated, proposals should not be exploitable.}\label{rec:attestation}

\section{Methodology}

\subsection{Research Questions}

Our main research objectives are (a) \emph{to shed light on the adoption of the Privacy Sandbox APIs during the full time period they were available in Chrome}, and (b) \emph{obtain insights and make suggestions about how and why the Privacy Sandbox might have failed}. We more specifically ask the following:

\begin{rqlist}[series=rquestion]
    \item What was the general adoption of these APIs over their availability period? \label{rq:adoption} (\autoref{sec:adoption})
    \item Who adopted the Privacy Sandbox APIs, which APIs did they use, and how? \label{rq:adopters} (\autoref{sec:adopters})
    \item Which supported APIs remain and why? \label{rq:left} (\autoref{sec:left})
\end{rqlist}

To answer these, we carry out a measurement analysis of the adoption of \emph{all} Privacy Sandbox web APIs during their \emph{entire} availability period (overview of findings in~\autoref{tab:findings-mapping}). This provides us with a systematic view into how the ecosystem reacted as a whole to the proposed changes.
This is in contrast with prior work (see~\autoref{sec:related-work}) that focused on just an individual proposal, a subset of APIs, and/or on a narrower time period, thus limiting prior insights to localized trends.

\subsection{Data Collection}

In order to obtain access to historical and longitudinal data on Privacy Sandbox adoption, we leverage all relevant publicly available datasets at our disposal. Mainly, we derive our results from the \textit{Chrome Telemetry} and \textit{HTTP Archive} datasets.
We also use version control information from a \textit{chromium} source tree component and from Google's \texttt{RWS} open-source repository to obtain the lists of some API adopters. Below, we describe each dataset and corresponding data collection methodology in more details, a discussion of related ethics considerations can be found in~\refappendix{appendix:ethics-considerations}.

\subsubsection{List of APIs} We start by collecting details on how to invoke each of the 13 identified proposals from~\autoref{tab:apis} by going through their respective explainer, developer documentation, and Chrome feature tracking status page. An overview of the permissions policy directives, JavaScript (JS) methods, HTML attributes, and HTTP headers detected in our measurement can be found in~\autoref{tab:appendix-apis-table} in Appendix.

\subsubsection{Blink Features} Google monitors the usage on the web of existing and new \textit{Blink} features by instructing Chrome to record in telemetry reports when one is used and detected in the wild. The \textit{Chrome Status} platform\footnote{\url{https://chromestatus.com/}} reports aggregated statistics such as daily usage of each feature of interest from Chrome users with telemetry turned on, across all platforms (Android, Windows, Mac, ChromeOS, and Linux) and channels (stable, beta, and dev)~\cite{googleChromiumDocsUseCounter2019}.

\noindent \textbf{\textit{Chrome Telemetry.}}
All APIs that we are measuring, except \texttt{RWS}, have one or several of their \textit{Blink} features monitored this way. \autoref{tab:overview-telemetry} presents an overview of the \textit{Chrome Telemetry} dataset we derive from these statistics for each API. Dates range from January 29, 2020 (i.e., towards the start of the initiative) to April 30, 2026, covering the full time period of the Privacy Sandbox initiative: from its start to after the deprecation of most of the APIs in January 2026, prior to conducting the analysis and submitting this paper. This offers us a unique perspective into the daily percentage of page loads across real users' browsing journeys where one of these APIs was actually used and not only implemented (\ref{rq:adoption} \&~\ref{rq:left}). In practice, in the rest of the paper, we average these daily percentages to report monthly or quarterly results, and combine all features under their parent API.

\begin{table}[!ht]
\footnotesize
\setlength\tabcolsep{1pt}
 \caption{Overview of the \textit{Chrome Telemetry} dataset.}
 \label{tab:overview-telemetry}
 \centering
 \begin{tabular}{ccccccc}
 \toprule
 & \multicolumn{3}{c}{\textbf{\scriptsize Number of Unique}} & \multicolumn{3}{c}{\textbf{\scriptsize Dates}}\\
 \textbf{\scriptsize API} & \textbf{\scriptsize Features} & \textbf{\scriptsize Entries} & \textbf{\scriptsize Days} & \textbf{\scriptsize Earliest} & &\textbf{\scriptsize Latest}\\
 \midrule
\texttt{\scriptsize Attribution Report.} &5&5\,167&2\,011&2020-10-02&&2026-04-30\\
\texttt{\scriptsize CHIPS} &1&1\,433&1\,433&2022-03-24&&2026-04-30\\
\texttt{\scriptsize FLoC} &1&656&656&2021-02-27&&2026-04-16\\
\texttt{\scriptsize FedCM} &17&10\,769&1\,470&2022-02-18&&2026-04-30\\
\texttt{\scriptsize Fenced Frames} &8&3\,938&1\,613&2021-08-01&&2026-04-30\\
\texttt{\scriptsize Private Aggregation} &6&6\,335&1\,241&2022-11-10&&2026-04-30\\
\texttt{\scriptsize Private State Tokens} &4&7\,936&2\,105&2020-06-19&&2026-04-30\\
\texttt{\scriptsize Protected Audience} &4&3\,029&1\,482&2022-03-31&&2026-04-30\\
\texttt{\scriptsize Shared Storage} &23&12\,558&1\,413&2022-06-17&&2026-04-30\\
\texttt{\scriptsize Storage Access} &30&14\,863&1\,922&2020-06-10&&2026-04-30\\
\texttt{\scriptsize Topics} &6&5\,333&1\,475&2022-04-04&&2026-04-30\\
\texttt{\scriptsize UA Client Hints} &30&46\,618&2\,192&2020-01-29&&2026-04-30\\
\midrule
\textbf{All} & 131 & 118\,635 & 2\,193 & 2020-01-29 & & 2026-04-30\\
 \bottomrule
 \end{tabular}
\end{table}

\subsubsection{HTTP Archive (HA)} Every month, the HA project\footnote{\url{https://httparchive.org/}} crawls millions of websites whose URLs come from the Chrome UX Report (CrUX) to record several web performance metrics (e.g., page loads and sizes, resources types, requests made, etc.)~\cite{HTTPArchive2010}. The results are made publicly available through a Google BigQuery database. These crawls are performed from Google Cloud locations in the United States through a Chrome instance on desktop and (emulated Android) mobile device. The crawler is non-interactive with the page visited (e.g., no click on any element or consent banner) and stateless (i.e., every visit is performed from a fresh Chrome instance with an empty cache and in a logged out state)~\cite{MethodologyWebAlmanac}. Thus, HA results usually represent a lower bound on the exact number of adopters and adoption rates over time, as more page interactivity could yield further APIs calls. 

\begin{table}[!ht]
\footnotesize
\setlength\tabcolsep{1pt}
 \caption{Overview of the \textit{HA-blink} dataset.}
 \label{tab:overview-ha-blink}
 \centering
 \begin{tabular}{cccccccc}
 \toprule
 & & \multicolumn{3}{c}{\textbf{\scriptsize Number of Unique}} & \multicolumn{3}{c}{\textbf{\scriptsize Crawls}}\\
 & \textbf{\scriptsize API} & \textbf{\scriptsize Features} & \textbf{\scriptsize Entries} & \textbf{\scriptsize Crawls} & \textbf{\scriptsize Earliest} & &\textbf{\scriptsize Latest}\\
 \midrule
\multirow{9}*{\rot{\textbf{Desktop}}} &\texttt{\scriptsize Attribution Reporting}& 3 & 50 & 32 & 2021-06 && 2026-04 \\
&\texttt{\scriptsize FedCM}& 9 & 202 & 41 & 2022-11 && 2026-04 \\
&\texttt{\scriptsize Fenced Frames}& 2 & 46 & 23 & 2024-06 && 2026-04 \\
&\texttt{\scriptsize Private State Tokens}& 2 & 63 & 58 & 2021-03 && 2026-04 \\
&\texttt{\scriptsize Protected Audience}& 3 & 49 & 23 & 2024-06 && 2026-04 \\
&\texttt{\scriptsize Shared Storage}& 10 & 164 & 23 & 2024-06 && 2026-04 \\
&\texttt{\scriptsize Storage Access}& 13 & 200 & 30 & 2023-11 && 2026-04 \\
&\texttt{\scriptsize Topics}& 4 & 87 & 23 & 2024-06 && 2026-04 \\
&\texttt{\scriptsize UA Client Hints}& 29 & 1\,282 & 60 & 2021-05 && 2026-04 \\
\midrule
& \textbf{All (desktop)} & 75 & 2\,143 & 62 & 2021-03 && 2026-04\\
 \bottomrule
 \end{tabular}
\end{table}

\noindent \textbf{\textit{HA-blink.}} Similar to \textit{Chrome Telemetry}, the HA crawler also detects if a \textit{Blink} feature has been used on a page. As a result, we derive the \textit{HA-blink} dataset that corresponds to the usage of Privacy Sandbox APIs features detected on the top 100k pages each month from both desktop and mobile devices. In practice, in this \textit{HA-blink} dataset we find evidence of the first use of a Privacy Sandbox API in March 2021, as shown by~\autoref{tab:overview-ha-blink}. This provides us with the prevalence of Privacy Sandbox features on top websites as a proxy for what real users encounter during their own browsing journey (\ref{rq:adoption} \&~\ref{rq:left}).

\noindent \textbf{\textit{HA-requests.}} To answer~\ref{rq:adopters} and~\ref{rq:left}, we not only want to know on which pages some APIs are being invoked, but also who is responsible for the call and how it is implemented. For that, we need to query the HTTP traffic data collected during HA crawls for the presence of one or several of the Privacy Sandbox APIs and for all different ways to use them (see~\autoref{tab:appendix-apis-table} in Appendix). As an example, for \texttt{Topics}, we check if \texttt{\small Sec-Browsing-Topics} is sent via an HTTP request header or if the \texttt{\small document.browsingTopics} JS method is called. Thus, we need to analyze HTTP response headers, but also check if the HTTP response itself (of type \texttt{\small script} or \texttt{\small html}) contains JS code implementing one of the APIs. 

A traditional problem is that this syntactic approach by string matching is usually error-prone, because, we can not distinguish whether a JS method was called or not (e.g., method name present in a branch not executed in practice or in a comment). Moreover, parsing naively all HA requests results since the existence of the Privacy Sandbox initiative would be very cost prohibitive for our analysis.

Here, our main insight is that we can leverage the prior detection of the \textit{Blink} Privacy Sandbox features to guide us on which requests from which pages should be further scrutinized. This helps us reduce query costs and control for incorrect detections (see discussion in~\refappendix{appendix:ethics-considerations}). Thus, we only run our syntactic matching on all root pages of the top 100k websites crawled since March 2021 from a desktop client onto which one of the feature in the \textit{HA-blink} dataset has been detected. We name the derived dataset \textit{HA-requests} and provide an overview of it in~\autoref{tab:parameters-ha-requests}. This dataset allows us to obtain more details such as which first- or third-party resources implemented one of the APIs. Thereafter, for readability, we combine all JS calls and HTTP headers under their parent API.

\begin{table}[!ht]
\footnotesize
\setlength\tabcolsep{1pt}
 \caption{Parameters and totals for the \textit{HA-requests} dataset.}
 \label{tab:parameters-ha-requests}
 \centering
 \begin{tabular}{ccc}
 \toprule
 &\textbf{Name} & \textbf{Value}\\
 \midrule
 \multirow{5}*{\rot{\textbf{Parameters}}} &Client & \texttt{desktop}\\
 &Date & 2021-03 to 2026-04\\
 &is\_root\_page & \texttt{true}\\
 &Rank & 100k\\
 &Type & \texttt{script} or \texttt{html}\\
 \midrule
 \multirow{6}*{\rot{\textbf{Total}}} & Pages parsed & 8\,403\,906\\
 &Responses headers parsed& 10\,645\,599\,709 ($\sim$1 TB in size)\\
 &Responses body parsed& 173\,959\,865 ($\sim$45 TB in size)\\
 &Pages with detection & 4\,325\,971\\
 &Responses headers with detection & 200\,274\,714\\
 &Responses body with detection& 18\,772\,657 \\
 \bottomrule
 \end{tabular}
\end{table}

\begin{figure*}[!ht]
  \includegraphics[width=\linewidth]{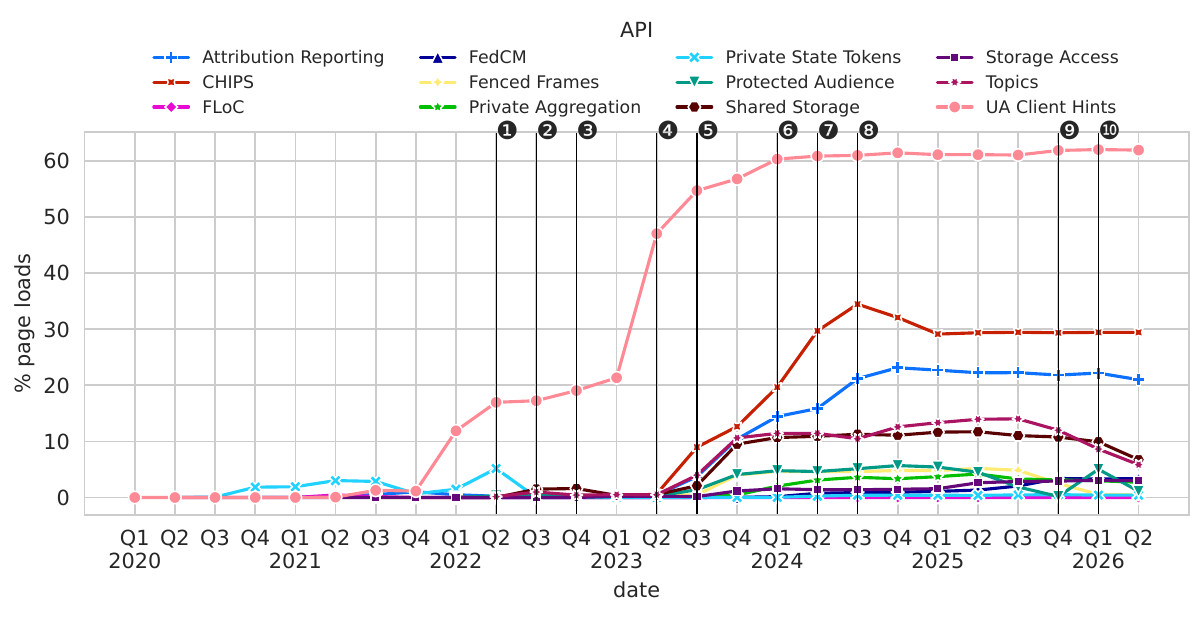}
  \caption{General adoption rates (quarterly percentage of page loads) over time from \textit{Chrome Telemetry} dataset (log-scale in~\autoref{fig:telemetry-log}). Vertical numbered lines correspond to major development events (see~\autoref{aside:timeline}).}
  \label{fig:telemetry}
\end{figure*}

\subsubsection{Open-source Repositories}
In addition to the \textit{HA} and \textit{Chrome Telemetry} datasets, we also leverage public version control information from a \textit{chromium} source tree component\footnote{\href{https://github.com/chromium/chromium/blob/87eacadcd3f978bdeb527f397c7b652435b0476a/components/privacy_sandbox/privacy_sandbox_attestations/}{privacy\_sandbox/privacy\_sandbox\_attestations}} and the \texttt{RWS} repository\footnote{\url{https://github.com/GoogleChrome/related-website-sets}}. This allows us to collect the longitudinal edits made to the attestation and \texttt{RWS} lists that were shipped with all Chrome instances (\ref{rq:adopters}).

\section{API Adoption (\ref{rq:adoption})}
\label{sec:adoption}

\noindent \textbf{\textit{Chrome Telemetry.}}
\autoref{fig:telemetry} shows the general evolution over time of the percentage of page loads onto which one of the features associated with one of the Privacy Sandbox APIs is detected. Note that this result on its own is the most complete report of the longitudinal adoption rates of the Privacy Sandbox APIs that we are aware of. From it, we can observe a very uneven situation as only 5 APIs were ever present on more than 11.7\% of page loads, and if we exclude them, it appears that adoption and usage by the ecosystem of Google's initiative remained very limited.

\finding{The adoption of Privacy Sandbox APIs is highly uneven with the most usage for \texttt{UA Client Hints}, \texttt{CHIPS}, \texttt{Attribution Reporting}, \texttt{Topics}, and \texttt{Shared Storage}, whereas the remaining APIs have seen very little to almost no use.}\label{finding:adoption-uneven}

While we discuss \texttt{CHIPS} and \texttt{UA Client Hints} in more details in~\autoref{sec:left}, it is worth mentioning here that the next 3 other APIs in number of usage have to do with web use cases that would have likely been directly the most impacted by TPC deprecation, i.e., ad targeting (\texttt{Topics}), conversion measurement (\texttt{Attribution Reporting}), and cross-site storage (\texttt{Shared Storage}). For the other APIs, their usage level may be indicative of a failure to convince on their functionality and usability beyond the specific tests carried out by third parties (see further discussion in~\autoref{sec:incentives-risks}).

\recommendation{Future attempts at web privacy-enhancing technologies should concentrate on the main use cases of advertising and conversion measurement.} \label{rec:api-use-cases-value} 

Regarding trends over time, we observe that adoption rates start increasing in real Chrome user page loads around the Q2 and Q3 of 2023. This is somewhat expected as this moment (see~\autoref{aside:timeline}) corresponds with Google announcing that they will start deprecating TPC for a small percentage of Chrome users starting early 2024, as well as with the general availability of the \texttt{Attribution Reporting}, \texttt{Fenced Frames}, \texttt{Private Aggregation}, \texttt{Protected Audience}, \texttt{Shared Storage}, and \texttt{Topics} APIs in Chrome 115. These two announcements combined have likely incentivized some actors to start experimenting with and implementing the APIs in preparation for the impact of TPC deprecation on their use cases, which was at that time still planned for the second half of 2024. However, we observe that this trend did not continue; by Q3 of 2024, adoption across the board has largely stabilized and reached about its maximum rates over the full time period of the Privacy Sandbox initiative. Indeed, in the coming months and quarters afterwards, we observe that rates have stagnated, with some APIs usage (e.g., \texttt{Topics}) already starting to decrease. Again, several factors could be taken into account here (see further discussion in~\autoref{sec:incentives-risks}), but Google's announcements with respect to the development of the initiative are strong indicators of the reasons why: in Q2 2024, TPC deprecation was postponed for a second time, before being canceled altogether in Q3 2024.
Finally, in Q4 2025, Google announced the deprecation of most of the Privacy Sandbox APIs for January 2026, explaining the adoption drop towards the end of our measurement period.

\noindent\textbf{\textit{HA-blink}.} For interested readers, we report in~\refappendix{sec:supplemental-figures} similar longitudinal adoption results that have been obtained on our other dataset \textit{HA-blink}.

\finding{Adoption rose after introduction of the APIs in 2023, but usage already peaked or stagnated by Q3 2024 when Chrome TPC deprecation was abandoned.}\label{finding:adoption-rise-plateau}

\recommendation{After this failed TPC deprecation in Chrome, new efforts will need to address the fact that the community will be reluctant to adopt future privacy proposals from Google and perhaps from others too.} \label{rec:advertising-apis}
\section{API Adopters (\ref{rq:adopters})}
\label{sec:adopters}

\begin{figure}[!ht]
  \includegraphics[width=\linewidth]{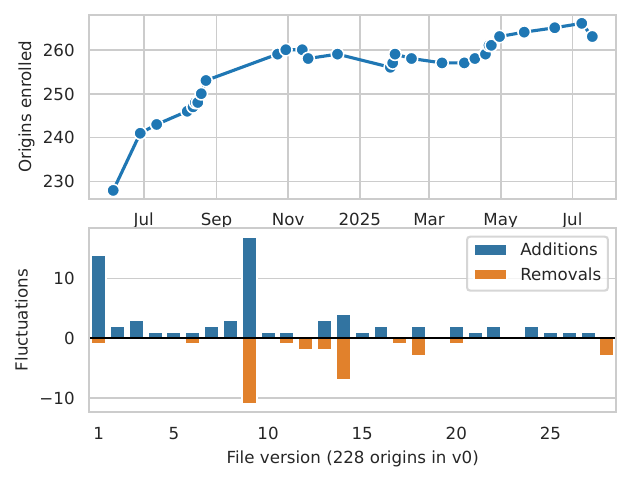}
  \caption{Number of enrolled origins from Jun'24 to Jul'25 and fluctuations between attestations file versions.}
  \label{fig:attestation-edits}
\end{figure}

\noindent\textbf{Attestation List.} \autoref{fig:attestation-edits} reports on the number of origins enrolled in the attestation process managed by Google. We observe that a total of 295 unique entities were at some point enrolled to be able to call the following 5 APIs: \texttt{Attribution Reporting}, \texttt{Private Aggregation}, \texttt{Protected Audience}, \texttt{Shared Storage}, and \texttt{Topics}. In practice, at any given point that number oscillated between 228 to 266 origins (mean of 256 origins). This indicates that a relatively few number of additions (i.e., newcomers) and removals (i.e., departures) occurred over time beyond the initial large set of first origins enrolled.

\begin{figure}[!ht]
    \includegraphics[width=\linewidth]{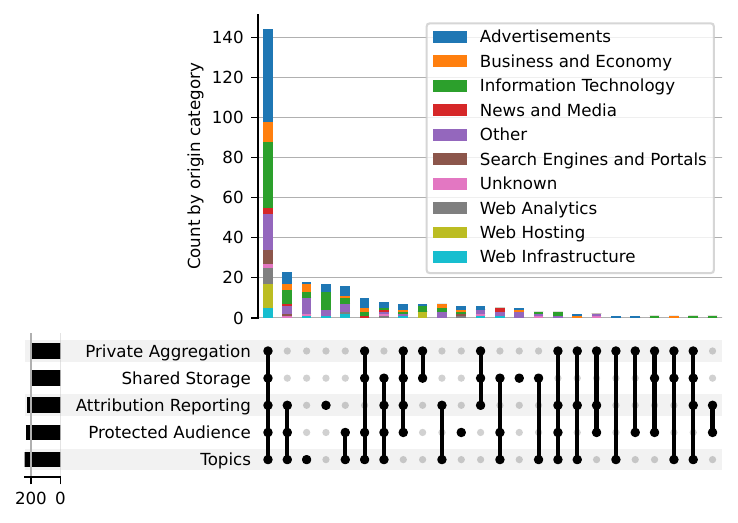}
  \caption{UpSet plot of APIs enrolled together according to attestations. Origin category is also reported.}
  \label{fig:attestation-apis}
\end{figure}

\autoref{fig:attestation-apis} shows an UpSet plot\footnote{An UpSet plot is a type of graph used to visualize overlap between multiple sets, as Venn diagrams scale poorly beyond 3 or 4 sets.} of all the different combinations of the 5 APIs that these 295 origins asked to enroll for. We also report on the categories of these origins as classified by \textit{Forcepoint}\footnote{\url{https://support.forcepoint.com/s/site-lookup}}. These categories indicate as expected that the most prominent adopters were advertisers or entities linked to some sort of online services, e.g., analytics, hosting, news, or search providers. Additionally, we observe that if the majority of origins enrolled to call all 5 APIs, others only requested specific combinations based on their needs. After the default 5 APIs mentioned previously, advertising-related use cases account for the top 4 of other requested combinations.

\noindent\textbf{\texttt{RWS} List.} Similarly, we go through all \texttt{RWS} submissions approved by Google from November 2022 to July 2025 on the corresponding GitHub repository. We find that a total of 73 sets made it to this reference list used by the Chrome browser. As a result, only 73 primary websites disclosed through \texttt{RWS} their relationships with other secondary services and websites they also own. By doing so, requests from and to these origins are not impacted by cross-site storage restrictions and partitioning from Chrome.

\begin{figure}[!ht]
\begin{subfigure}{\linewidth}
    \centering
    \includegraphics[width=\linewidth]{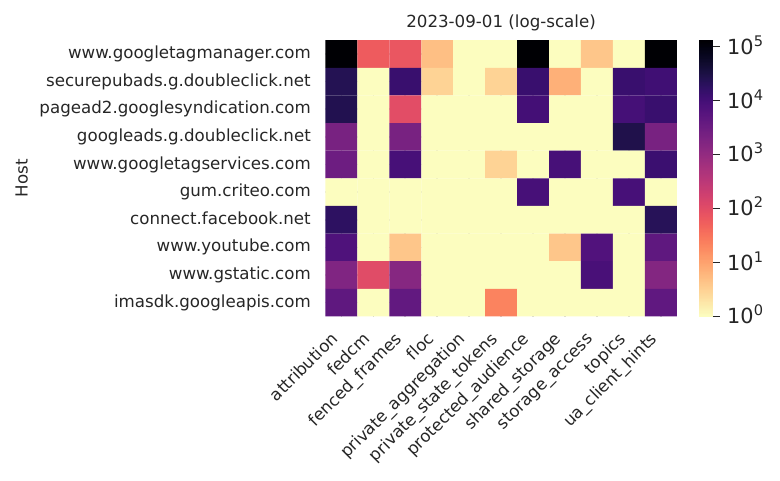}
    \caption{September 2023}
    \label{fig:adopters-2023}
\end{subfigure}
\begin{subfigure}{\linewidth}
    \centering
    \includegraphics[width=\linewidth]{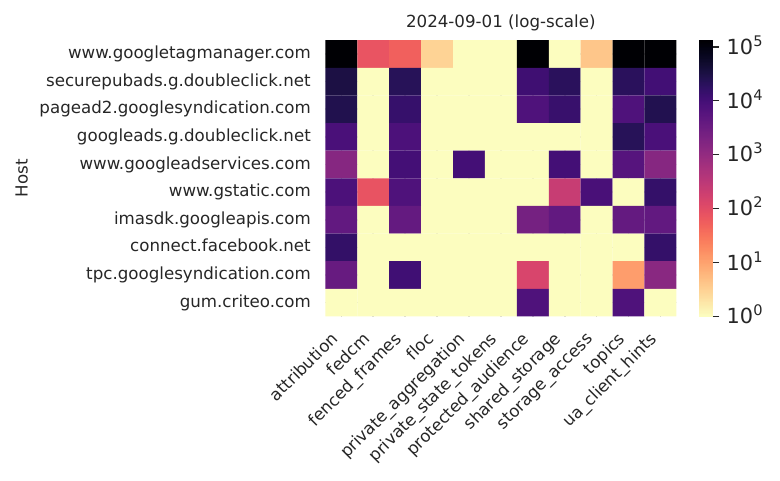}
    \caption{September 2024}
    \label{fig:adopters-2024}
\end{subfigure}
\begin{subfigure}{\linewidth}
    \centering
    \includegraphics[width=\linewidth]{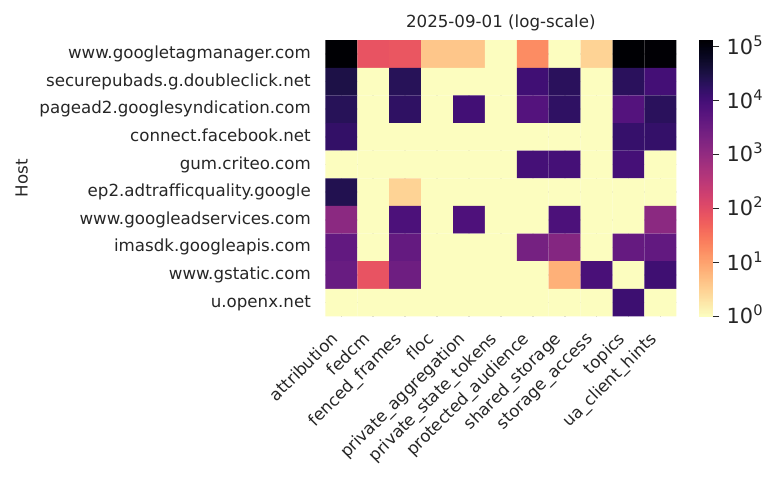}
    \caption{September 2025}
    \label{fig:adopters-2025}
\end{subfigure}
\caption{API usage (log-scale) for each of the top 10 API adopters ranked according to their combined total use. Data obtained from \textit{HA-requests} dataset for 3 specific dates, one year apart each.}
\label{fig:adopters}
\end{figure}

\noindent\textbf{\textit{HA-requests.}} Next, we pivot to measuring the JS implementation of one of the Privacy Sandbox APIs by web actors in the HA monthly crawl. \autoref{fig:adopters} provides the top 10 API adopters (hosts) we detect during a respective HA monthly crawl. These heatmaps showcase the total number of implementations we detect for each API by this specific host across all visited pages. Hosts are ranked per their combined total API usage.
We plot heatmaps for 3 specific dates spaced apart from each other by a year. These specific dates were picked as they correspond to the early availability of APIs (September 2023), the time when adoption peaked (September 2024), and before Google announced the end of the initiative (September 2025). 
Comparing hosts across these 3 dates, we find that the set of top adopters is somewhat consistent over the years meaning that testers driving adoption at early stages of the project were still the main ones responsible for API usage at later ones. 

\finding{The majority of adopters were active starting in the early days of the initiative.}\label{finding:adopters-attestation}

\recommendation{The value of proposals should be demonstrated upfront to secure long term commitment from early adopters and encourage further adoption by others.}\label{rec:early-testers}

\section{Supported APIs (\ref{rq:left})}
\label{sec:left}

As a reminder, only 6 of the Privacy Sandbox APIs will continue to be supported in Chrome: \texttt{CHIPS}, \texttt{FedCM}, \texttt{Fenced Frames}, \texttt{Private State Tokens}, \texttt{Storage Access}, and \texttt{UA Client Hints}.

\begin{table}[!ht]
\centering
\caption{Max and recent adoption for supported APIs for \textit{Chrome Telemetry} dataset (\% page loads).}
\label{tab:remaining-apis}
  \begin{tabular}{ccccc}
  \toprule
    \textbf{API} & \textbf{Max} & \textbf{Q3'25} & \textbf{Q4'25} & \textbf{Q1'26}\\
    \midrule
    \texttt{\scriptsize CHIPS} & 34.45 &  29.44 &  29.36 &  29.43\\
    \texttt{\scriptsize FedCM} & 3.39 &  2.04 &  3.39 &  3.28\\
    \texttt{\scriptsize Fenced Frames} & 5.19  &  4.84 &  2.51 &  0.53\\
    \texttt{\scriptsize Private State Tokens} & 5.16 &  0.46 &  0.50 &  0.42\\
    \texttt{\scriptsize Storage Access} & 3.05 &  2.85 &  2.92 &  3.05\\
    \texttt{\scriptsize UA Client Hints} & 62.01 &  61.04 &  61.86 &  62.01\\
    \bottomrule
  \end{tabular}
\end{table}

\begin{figure}[!ht]
\centering
    \includegraphics[width=\linewidth]{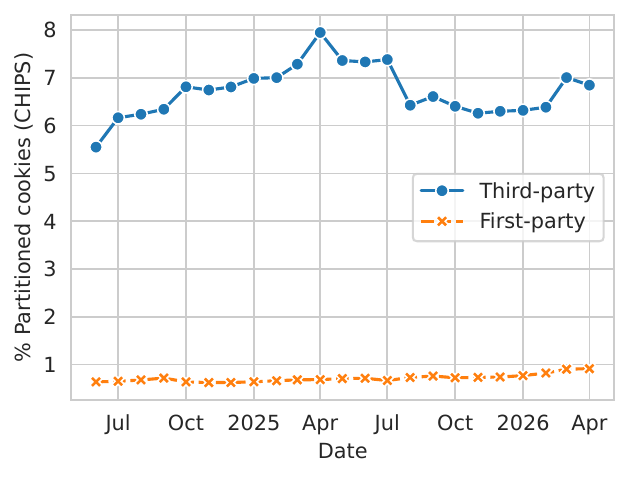}
  \caption{Partitioned cookies (\texttt{CHIPS}) from Jun'24 to Apr'26 among first- and third-party cookies (\textit{HA-requests}).}
  \label{fig:chips-percentage}
\end{figure}

\begin{table}[!ht]
\footnotesize
\setlength\tabcolsep{1pt}
 \caption{Top partitioned TPC set on pages with \texttt{CHIPS} cookies for April 2026 (\textit{HA-requests}).}
 \label{tab:chips-setter}
 \centering
 \begin{tabular}{ccc}
 \toprule
 \textbf{Rank}&\textbf{Cookie (name - domain)}&\textbf{\%}\\
 \midrule
1& cto\_bundle - .criteo.com & 43.91\\
2& audit\_p - .rubiconproject.com & 34.32\\
3& khaos\_p - .rubiconproject.com & 34.32\\
4& receive-cookie-deprecation - .rubiconproject.com & 29.88\\
5& ts - .creativecdn.com & 29.02\\
6& VP - .contextweb.com & 22.15\\
7& pamuid2 - .a-mo.net & 21.09\\
8& psd\_amuid2 - .sync.a-mo.net & 20.28\\
9& pb\_rtb\_ev\_part - .contextweb.com & 19.93\\
10& server\_tracking\_bdsp\_uid - .tracookiepixel.xyz & 18.88\\

 \bottomrule
 \end{tabular}
\end{table}

\noindent \textbf{\textit{Chrome Telemetry} \& \textit{HA-requests}.}
 \autoref{tab:remaining-apis} shows the maximum and recent quarterly adoption rates of these APIs pre- and post- Google's deprecation announcement of late 2025.
We observe that \texttt{UA Client Hints} is the API that is used the most predominantly, i.e., on more than 62\% of pages, and that all other 5 APIs currently exhibit a much lower adoption rate. 
Indeed, \texttt{CHIPS} is present on about 34.5\% of page loads, which could be interpreted as a somewhat high usage. However, in practice an average of only 6.7\% of all TPC set during each crawl since June 2024 are partitioned (max of 7.9\% of TPC) as shown by~\autoref{fig:chips-percentage}.
We report in~\autoref{tab:chips-setter} the top 10 partitioned TPC set on pages with \texttt{CHIPS} cookies. We find that programmatic advertising and digital ad tech companies---embedded on a large number of websites---are responsible for setting the majority of these partitioned TPC.

\finding{Except for \texttt{UA Client Hints}, the other APIs still supported by Chrome exhibit low usage.}\label{finding:remaining-apis}

\finding{\texttt{CHIPS} adoption is mainly driven by a few actors with large coverage of the web.}\label{finding:remaining-chips}

\recommendation{Future attempts may require additional external incentives such as pressure stemming from regulatory requirements, broader users' awareness, or privacy protections shipped by competing browsers.}\label{rec:external-incentives}
 
\begin{table*}[!ht]
\footnotesize
\setlength\tabcolsep{1pt}
 \caption{Top hosts and corresponding scripts implementing \texttt{UA Client Hints} through JS in April 2026 (\textit{HA-requests}).}
 \label{tab:ua-script}
 \centering
 \begin{tabular}{ccccc}
 \toprule
 \textbf{Rank}&\textbf{Host}&\textbf{Script}&\textbf{\% Pages}&\textbf{Use Case}\\
 \midrule
1 & www.googletagmanager.com & js & 68.23 & Google tracking tag management \\
2 & www.googletagmanager.com & gtm.js & 40.24 & Google tracking tag management\\
3 & connect.facebook.net & fbevents.js & 23.78 & Meta tracking pixel \\
4 & securepubads.g.doubleclick.net & gpt.js & 15.11 & Google ad serving library\\
5 & pagead2.googlesyndication.com & ufs\_web\_display.js & 12.29 & Google ad serving library\\
6 & scripts.clarity.ms & clarity.js & 12.28 & Microsoft user behavior analytics\\
7 & www.googletagmanager.com & destination & 11.62 &Google tracking tag configuration\\
8 & pagead2.googlesyndication.com & adsbygoogle.js & 11.44 & Google ad serving library\\
9 & bat.bing.com & bat.js & 9.56 & Microsoft (Bing) tracking and ad serving\\
10 & www.gstatic.com & recaptcha\_\_en.js & 6.59 & Google reCAPTCHA service\\
 \bottomrule
 \end{tabular}
\end{table*}

Perhaps, the case of \texttt{UA Client Hints} illustrates best the dynamics of the ecosystem. While access to \texttt{UA} high entropy features have been restricted by default on every major browser, Chrome restored access to them by allowing third-parties to just call the new \texttt{UA Client Hints} API. 
\autoref{tab:ua-script} shows the top 10 hosts and scripts that implement the \texttt{UA Client Hints} API through JS in April 2026 in the \textit{HA-requests} dataset per pages percentage. We also annotated the list with the main use case for each of these scripts. We observe that the wide implementation of \texttt{UA Client Hints} we detect in \textit{HA-requests} is attributable to the tracking, advertising, analytics, and bot and spam protection services provided by Google, Meta, and Microsoft. 
This continuing use of \texttt{UA Client Hints} by such tracking entities to access higher entropy hints raises concerns. Indeed, \texttt{UA} information was traditionally used to fingerprint users before \texttt{UA} reduction efforts from browsers. It appears that trackers restarted doing so through the \texttt{UA Client Hints} API~\cite{senolUnveilingImpactUserAgent2023,ilgazInvestigatingHighEntropyClient2025}. Moreover, a prior measurement study from Intumwayase et al., showed that altering the \texttt{UA} does not result in visually different browsing experience, suggesting that this information is not needed anymore for the functioning of the modern web~\cite{intumwayaseUARadarExploringImpact2023}.

\finding{The \texttt{UA Client Hints} API is being abused for fingerprinting as it restores access to \texttt{UA} features that have been restricted by default otherwise.}\label{finding:ua-ch-abuse}

\recommendation{The continuing use of \texttt{UA Client Hints} should be justified and user-friendly controls to disable the API must be provided to users by Chrome and other browsers supporting the API.}\label{rec:apis-value}

\section{Discussion}
\label{sec:discussion}

Next, we reflect on the Privacy Sandbox experiment and what we learned through our measurement findings. We also discuss the need for user consent with respect to Google's initiative and contextualize how the Privacy Sandbox fits into a broader narrative of events and tracking mitigation implemented by other web browsers. While doing so, we also offer some suggestions for future work.

\subsection{Reflections}
\label{sec:incentives-risks}

\subsubsection{Findings}
Through our study, we observed that some actors in the ecosystem made a significant attempt towards testing and adopting these new Privacy Sandbox APIs. They did so for an extended period of time based on the visible signals in our datasets. This resulted in a very uneven situation with some proposals that saw substantial usage on the web (mainly for the advertising-related mechanisms), while others not nearly as much. Additionally, adoption quickly rose after the APIs' introduction, but reached its maximum and started to stagnate around the time Google postponed and then canceled TPC deprecation, after which overall usage never appears to have gone up again. While we cannot be certain about the specific reasons from actors for experimenting or not with these APIs, without being them or Google, we provide thereafter a few possible interpretations.

\subsubsection{Incentives and Risks}
Ultimately, actors had to contemplate several incentives and risk factors when making their decisions to partially try, fully adopt, or not use these mechanisms. A potential reason is that actors were just trying to get ahead of the TPC deprecation and associated impact on their business models, but as soon as Google stopped driving the initiative, their interests plummeted too. Another possibility is that this represented a real opportunity for some to adapt their practices and figure out ways forward that more adequately respect users' privacy. Let's also not omit that testing and implementing these proposals into live systems and products represents an important commitment in terms of time, resources, people, money, etc., that not everyone was likely able to take on. Additionally, during its development, some aspects of the Privacy Sandbox initiative were still actively evolving, unresolved, or unclear, such as the privacy and utility trade-offs these actors were agreeing to, potential privacy issues and violations, and legal exposure from using these new mechanisms (see~\autoref{sec:user-consent}). All of these factors, and others, could have impacted the decision-making of actors to adopt or not adopt the Privacy Sandbox APIs.

\subsubsection{Nuances}
At the same time, we have seen how the continuing use of \texttt{UA Client Hints} weakens \texttt{UA} reduction efforts in Chrome and discussed that its usage should be re-evaluated given its abuse in browser fingerprinting.

\subsubsection{Takeaway}
Overall, our data suggest that a nontrivial portion of actors from the advertising ecosystem are willing to experiment with web privacy proposals and that there is a real market for such approaches, whether Privacy Sandbox was the right solution or not. As such, future attempts at privacy-enhancing technologies for the web may have a better chance if economics or incentives change, for instance through pressure stemming from new regulatory requirements and broader awareness from users (\autoref{sec:user-consent}) or privacy mitigations shipped by other browsers (\autoref{sec:other-browsers}).

\subsection{Privacy Sandbox \& User Consent}
\label{sec:user-consent}
\subsubsection{Regulation Compliance}
Although the Privacy Sandbox aimed at deprecating TPC in Chrome, it is likely that the need for consent management systems and banners on the web would have remained. Indeed, similar to other web technologies, the use of the Privacy Sandbox proposals by web actors would have still required user consent in some jurisdictions.
In fact, Google clearly stated in their compliance FAQ that the \emph{``use of each of the Privacy Sandbox APIs involves accessing data that is stored on the user's device''}~\cite{googlePrivacyrelatedComplianceFAQs2025} and \emph{``strictly necessary''} exemptions\footnote{This is a direct reference to the EU's ePrivacy Directive~\cite{Directive2002582002,Directive20091362009}.} to explicit consent would only apply to very narrow cases and specific APIs, such as account protection (\texttt{FedCM}) or fraud prevention (\texttt{Private State Tokens}).
Anecdotally, Google, itself, specifically prompted Chrome users located in the European Economic Area and the UK\footnote{Despite UK's exit from the EU, UK's privacy laws are still very much aligned with the EU's ePrivacy Directive and GDPR.} for approval before rolling out \texttt{Topics} to them in 2023.

\subsubsection{Continuing Need for Consent Mechanism Systems} As such, the proposed TPC deprecation in Chrome would not have addressed the continuing need to repetitively ask users for consent during their browsing journey. Similarly, Google's Privacy Sandbox initiative did not appear to consider streamlining users' experience with respect to consent until only July 2024. Then, Google confirmed that TPC deprecation was not considered anymore in Chrome and announced that a new standalone prompt was being planned instead to let \textit{``people make an informed choice that applies across their web browsing''}. However, this mechanism was never deployed~\cite{chavezNewPathPrivacy2024,chavezNextStepsPrivacy2025}.
In retrospect, this could be interpreted as a missed opportunity. Indeed, through its Privacy Sandbox proposals, Google already acknowledged and proposed the idea of relying more heavily on the browser to mediate access to the newly introduced resources. So, perhaps, the browser could also enforce that users' consent was actually provided before setting cookies for instance, and if needed technically apply users' choices on their behalf. This would require browsers to assume more predominantly their role of gatekeeper.

\subsubsection{Potential Privacy Violations} As a result, the use of some Privacy Sandbox APIs without explicit user consent and agreement may constitute---under some jurisdictions---a violation of privacy laws.
Given that our results have been obtained from crawls whose instrumentation does not interact with visited websites nor click through banners, it appears that the API calls observed in HA data through \textit{blink features} were largely made without such consent being given to these web actors.
Verna et al., performed a consent-aware crawl on \texttt{Topics} and found adopters ignoring consent and using the API anyway~\cite{vernaUnderstandingTopicsAPI2025}. A similar analysis of all proposals would help determine if APIs are indeed being invoked before or after the acceptance or denial of consent mechanisms. Unfortunately, no publicly available dataset has such pre- and post-consent information. Given the deprecation of most APIs early 2026, we focused our efforts instead on reporting on a missing aspect in the field, i.e., a longitudinal adoption measurement of all these APIs.

\recommendation{User consent and legal implications should be integral considerations at API design. For instance, browsers should be able to enforce user consent before returning results rather than relying on non abuse.} \label{rec:consent}

\subsection{Privacy Protections Across Browsers}
\label{sec:other-browsers}
\subsubsection{Third-party Cookies Restrictions}
While Google did not deprecate TPC in Chrome with the Privacy Sandbox, some new cookie management policies were implemented such as the default changes to the \texttt{SameSite} cookie attribute~\cite{googleDevelopersGetReady2019}, as well as the opt-in possibility for third-parties to use the \texttt{Storage Access} API~\cite{mdnStorageAccessAPI2024} and \texttt{CHIPS} partitioned cookies.
In the meantime, other web browsers like Brave, Firefox, and Safari already had or enabled by default several TPC restrictions, as described below.

\noindent \textbf{Brave} applies, perhaps, the most aggressive policy by blocking all third-party application storage (e.g., cookies, \texttt{localStorage}, \texttt{indexedDB}). Since 2021, to reduce website breakage, third parties can only access an ephemeral and partitioned storage for the duration of the session~\cite{braveEphemeralThirdpartySite2021}.

\noindent \textbf{Firefox} has been blocking TPC set by known trackers since 2019 with \emph{Enhanced Tracking Protection}. In 2022, Firefox rolled out \emph{Total Cookie Protection} to partition in different cookie jars all TPC per the first-party origin visited~\cite{mozillaFirefoxRollsOut2022}.

\noindent \textbf{Safari} blocks all TPC by default, but has a few exceptions in place if \texttt{Storage Access} is being used or the third-party domain was visited as first-party by the user in the last 30 days. Additionally, an ML model detects domains with third-party storage access that perform web tracking~\cite{TrackingPreventionWebKit2020}.

\subsubsection{Other Privacy Features Shipped in the Meantime} 
In parallel to Google's initiative, the following privacy solutions were made generally available by other browsers.

\noindent \textbf{Brave} proposed \emph{Farbling} in 2020 as a redesign of its existing fingerprinting defenses; now by default Brave is randomly perturbing the output of browser APIs used by fingerprinting scripts (e.g., \texttt{Canvas}, \texttt{WebGL}, \texttt{Audio}, \texttt{Plugins}, \texttt{UA}, \texttt{EnumerateDevices}, and \texttt{HardwareConcurrency}). This is done in a way that still allows legitimate use of these features by regular  websites~\cite{braveFingerprintingDefenses202020,braveFingerprintRandomization2020}.

\noindent \textbf{Firefox} made modifications to how client-side state is managed through implementing network, state, and dynamic partitioning in 2021 and 2022, followed by more fingerprinting protections adding noise to the \texttt{Canvas} API in 2023~\cite{mdnStatePartitioningPrivacy2024}.

\noindent \textbf{Safari} rolled out \textit{Advanced Fingerprinting Protection} to private mode in 2024 and made it generally available by default in 2025. These defenses include filtering of tracking parameters in URLs, blocking network requests from known trackers, hiding IP addresses, fixing values, and injecting noise into APIs traditionally used by fingerprinters~\cite{wilanderPrivateBrowsing202024}.
Note how in contrast to the opt-in mitigations now in Chrome, these solutions have been enabled by default on other browsers, thus enhancing privacy for all of their users.

\recommendation{Vetted and meaningful privacy protections should be rolled out by default to web users.} \label{rec:privacy-by-default}

\subsection{Adoption of the Android Proposals}
We focus on the web APIs of the Privacy Sandbox, leaving to future work a potential extension of the measurement to the Android proposals. Nonetheless, we give some pointers to researchers interested in such endeavors. We suggest using a dataset of Android applications such as \texttt{AndroZoo}~\cite{androzoo2016} to access archival versions of different smartphone applications and create an \emph{.apk} decompiler to detect the potential use of the Android APIs. This could be complemented with scanning popular open-source Android apps or libraries, as the source code and historical versions are readily available.

\begin{table*}[!ht]
  \centering
\caption{Comparison of prior work on the measurement adoption of Privacy Sandbox APIs.}
\label{tab:prior-work}
\setlength\tabcolsep{5pt}
\begin{tabular}{@{}ccccccccccccccccc@{}}

\toprule
\textbf{Work} & \rot{Attestation}& \rot{\texttt{Attr. Reporting}} & \rot{\texttt{CHIPS}} & \rot{\texttt{FedCM}} & \rot{\texttt{FLoC}} & \rot{\texttt{Fenced Frames}} & \rot{\texttt{Private Aggregation}} & \rot{\texttt{Priv. State Tokens}} & \rot{\texttt{Protected Audience}} & \rot{\texttt{RWS}} & \rot{\texttt{Shared Storage}} & \rot{\texttt{Storage Access}} & \rot{\texttt{Topics}} & \rot{\texttt{UA Client Hints}}  & \textbf{Vantage Point (Timeline)} &\rot{\textbf{Open Data}} \\ \midrule

RTB House~\cite{ruminskiFindingsEarlyFledge2022}&\emptypie{}&\emptypie{}&\emptypie{}&\emptypie{}&\emptypie{}&\emptypie{}&\emptypie{}&\emptypie{}&\pie{360}&\emptypie{}&\emptypie{}&\emptypie{}&\emptypie{}&\emptypie{}&DSP access (Apr-Aug'21)&\xmark\\

Senol et al.~\cite{senolUnveilingImpactUserAgent2023}&\emptypie{}&\emptypie{}&\emptypie{}&\emptypie{}&\emptypie{}&\emptypie{}&\emptypie{}&\emptypie{}&\emptypie{}&\emptypie{}&\emptypie{}&\emptypie{}&\emptypie{}&\pie{360}&CrUX 100k (Jun'23)&\checkmark\\ 

Wiefling et al.~\cite{wieflingPrivacyMeasureTurned2024}&\emptypie{}&\emptypie{}&\emptypie{}&\emptypie{}&\emptypie{}&\emptypie{}&\emptypie{}&\emptypie{}&\emptypie{}&\emptypie{}&\emptypie{}&\emptypie{}&\emptypie{}&\pie{360}&\begin{tabular}[c]{@{}c@{}}HA top 8M (Jul'17-Dec'23) +\\ 327k login pages (Aug'22-Dec'23)\end{tabular} &$\checkmark^\ast$\\ 

Calderonio et al.~\cite{calderonioFledgingWillContinue2024}&\emptypie{}&\emptypie{}&\emptypie{}&\emptypie{}&\emptypie{}&\emptypie{}&\emptypie{}&\emptypie{}&\pie{-180}&\emptypie{}&\emptypie{}&\emptypie{}&\emptypie{}&\emptypie{}&Tranco 70k (Jun-Sep'23)&\xmark\\

Philipse et al.~\cite{philipsePostThirdPartyCookies2024}&\emptypie{}&\emptypie{}&\emptypie{}&\emptypie{}&\emptypie{}&\emptypie{}&\emptypie{}&\emptypie{}&\pie{-180}&\emptypie{}&\emptypie{}&\emptypie{}&\pie{-180}&\emptypie{}&Tranco 10k (Dec'23)&\checkmark\\

McQuistin et al.~\cite{mcquistinFirstLookRelated2024}&\emptypie{}&\emptypie{}&\emptypie{}&\emptypie{}&\emptypie{}&\emptypie{}&\emptypie{}&\emptypie{}&\emptypie{}&\pie{360}&\emptypie{}&\emptypie{}&\emptypie{}&\emptypie{}&\texttt{RWS} edits (Nov'22-Mar'24) &\checkmark\\

Zöllner et al.~\cite{zollnerFirstLookCookies2025a}&\emptypie{}&\emptypie{}&\pie{360}&\emptypie{}&\emptypie{}&\emptypie{}&\emptypie{}&\emptypie{}&\emptypie{}&\emptypie{}&\emptypie{}&\emptypie{}&\emptypie{}&\emptypie{}&HA top 8M (May'22-Sep'24)&\checkmark\\

Johnson et al.~\cite{johnsonAdventPrivacycentricDigital2024,johnsonUnearthingPrivacyEnhancingAd2024}&\emptypie{}&\emptypie{}&\emptypie{}&\emptypie{}&\emptypie{}&\emptypie{}&\emptypie{}&\emptypie{}&\pie{-180}&\pie{360}&\emptypie{}&\emptypie{}&\pie{-180}&\emptypie{}&Sincera access (May'23-Jun'25)&\xmark\\ 

Web Almanac~\cite{beuginWebAlmanacCookies2024,vekariaWebAlmanacPrivacy2024,beuginWebAlmanacCookies2026}&\pie{360}&\pie{-180}&\pie{360}&\emptypie{}&\pie{90}&\emptypie{}&\emptypie{}&\emptypie{}&\pie{90}&\pie{360}&\emptypie{}&\emptypie{}&\pie{90}&\pie{90}&HA top 1M (Jun'24 + Jul'25)&\checkmark\\

Verna et al.~\cite{vernaUnderstandingTopicsAPI2025}&\pie{90}&\emptypie{}&\emptypie{}&\emptypie{}&\emptypie{}&\emptypie{}&\emptypie{}&\emptypie{}&\emptypie{}&\emptypie{}&\emptypie{}&\emptypie{}&\pie{360}&\emptypie{}&Tranco 50k (Mar'24 + Aug'24-Mar'25)&\checkmark\\

Ilgaz et al.~\cite{ilgazInvestigatingHighEntropyClient2025}&\emptypie{}&\emptypie{}&\emptypie{}&\emptypie{}&\emptypie{}&\emptypie{}&\emptypie{}&\emptypie{}&\emptypie{}&\emptypie{}&\emptypie{}&\emptypie{}&\emptypie{}&\pie{360}&Tranco 100k (Apr'25)&\checkmark\\

Grib et al.~\cite{gribRiseFallGoogle2026} (concurrent work)&\pie{90}&\pie{360}&\pie{360}&\pie{-180}&\emptypie{}&\pie{-180}&\pie{360}&\pie{-180}&\pie{-180}&\emptypie{}&\pie{-180}&\pie{-180}&\pie{360}&\emptypie{}&Tranco 10k (Jun'25-Jan'26)&\checkmark\\
\midrule

Ours&\pie{360}&\pie{360}&\pie{360}&\pie{360}&\pie{360}&\pie{360}&\pie{360}&\pie{360}&\pie{360}&\pie{360}&\pie{360}&\pie{360}&\pie{360}&\pie{360}&\begin{tabular}[c]{@{}c@{}}HA top \NumTopWebsites (Mar'21-Apr'26) +\\Chrome Telemetry (Jan'20-Apr'26) +\\Attestation edits (Jun'24-Jul'25) +\\\texttt{RWS} edits (Nov'22-Jul'25)\end{tabular}&\checkmark\\

\bottomrule
\end{tabular}\\
\pie{90}: Limited analysis.
\hspace{1em}
\pie{-180}: Partial instrumentation only, e.g., JS API(s) and/or HTTP header(s) missing.
\hspace{1em}
$\checkmark^\ast$: Artifact only partially available.

\end{table*}

\section{Related Work on Measurement}
\label{sec:related-work}

The analyses on the Privacy Sandbox can be broadly divided into two categories: (a) privacy evaluations of the guarantees claimed by some proposals (see~\autoref{sec:related-work-privacy}), and (b) measurement studies of an individual proposal or a subset of them (described thereafter).
In comparison to our approach's scope, the following measurements were on limited APIs, narrow time periods, specific invocation methods, or a combination thereof (refer to overview in~\autoref{tab:prior-work}). Instead, we systematically cover all features that were made available in Chrome, monitor their adoption longitudinally thanks to historical data (5+ years), on the top \NumTopWebsites websites (per CrUX rank) and as experienced by Chrome users when browsing.

\subsection{Measurement of Individual Proposals}

\noindent\textbf{\texttt{Protected Audience.}}
In 2022, the demand-side platform (DSP) RTB House published a whitepaper sharing some findings on their early experiments performed on \texttt{Protected Audience} through their integrations with some of their partners, i.e., supply-side platforms (SSP) and publishers~\cite{ruminskiFindingsEarlyFledge2022}. Between April and August 2022, RTB House reports having added 1.2M users to interest groups around the world and made over 7M \texttt{Protected Audience} impressions. Overall, RTB House noticed a low engagement from the advertising industry; observing only Google and Criteo performing other tests, SSPs prioritizing other integrations or waiting for more clarity on how auctions will be executed, and publishers being largely unaware of the Privacy Sandbox origin trials.
Calderonio et al., studied the \texttt{Protected Audience} ecosystem just around its official deployment by crawling the Tranco top 70k websites from June to September 2023~\cite{calderonioFledgingWillContinue2024}, and found that the majority of the ad auctions through the API were run by Google.
A result observed as well by Philipse et al., in their own crawl of the Tranco 10k websites performed in December 2023 from an EU (Amsterdam) and US (New York City) location~\cite{philipsePostThirdPartyCookies2024}. In addition to confirming the prevalence of Google's DoubleClick in the calls to the \texttt{Protected Audience}, they also found 6 other advertising companies, and exposed and analyzed their bidding logic. They also observed as much as 3 times more calls to \texttt{Topics} than \texttt{Protected Audience}, and overall much fewer requests to join interest groups when crawling from the EU than the US, likely due to the simplicity of calling \texttt{Topics} and to EU consent regulations, respectively.

\noindent\textbf{\texttt{UA Client Hints.}}
In June 2023, Senol et al., quantified that high-entropy \texttt{UA Client Hints} were accessed on about 60\% of the home pages of the top 100k websites ~\cite{senolUnveilingImpactUserAgent2023}. They found that a majority of these calls were made by trackers and advertisers through JS (limited use of HTTP headers) and that the API results were routinely exfiltrated to remote servers.
In a different measurement, Wiefling et al., looked into the use of \texttt{Client Hints} headers, an earlier mechanism introduced in 2013 from which \texttt{UA Client Hints} is the most widely implemented category. In this study, Wiefling et al., queried HA data from July 2017 to December 2023 and crawled 327k login pages monthly from August 2022 to December 2023, also finding concerning access rates to \texttt{UA Client Hints} in third-party contexts~\cite{wieflingPrivacyMeasureTurned2024}.
In a followup, Ilgaz et al., complemented the prior studies on \texttt{Client Hints} headers by investigating their usage in HTTP/2 and HTTP/3 on a crawl of the top 100k websites in April 2025 from an EU (Amsterdam) and US (New York City) location~\cite{ilgazInvestigatingHighEntropyClient2025}. The reported results show a five to ten-fold growth in adoption rates in 2025 compared to 2023 measurements.

\noindent\textbf{\texttt{RWS.}} McQuistin et al., carried out a user study showing that users failed to correctly identify that a relationship often exists between websites of a same \texttt{RWS}~\cite{mcquistinFirstLookRelated2024}. Additionally, they also parsed and reported on the pull requests from November 2022 to March 2024 on the GitHub \texttt{RWS} repository that Google used to manage edits to the \texttt{RWS} list.

\noindent\textbf{\texttt{CHIPS.}} Zöllner et al., studied the adoption of partitioned cookies among the HA data from May 2022 to September 2024~\cite{zollnerFirstLookCookies2025a}. Their analysis showcased an increase in adoption of \texttt{CHIPS} around the start of the TPC deprecation test in Chrome, and a slow-down even before the announcement that TPC would still be supported. They also found asymmetric choices being made by adopters on enabling \texttt{CHIPS} by default or only on certain pages and sites, and pointed out to mistakes in the implementation by some developers of the new default requirements for the \texttt{SameSite} attribute.

\noindent\textbf{\texttt{Topics.}} Verna et al., measured the prevalence of the \texttt{Topics} API on the top 50k websites weekly from August 2024 to March 2025~\cite{vernaUnderstandingTopicsAPI2025}. Their crawler being consent-aware recorded if the API was used before or after consent was given and from different locations using a VPN. They found that \texttt{Topics} was quite often triggered before any consent to user cookies or similar tracking technologies was granted, which could be interpreted as a privacy violation under some regulations such as GDPR. They also briefly discussed the need to have an attestation to call some specific APIs.

\subsection{Measurement of a Subset of Proposals} 

\noindent\textbf{Sincera.} Johnson et al., partnered with Sincera, an advertising telemetry service, to obtain data and release an online dashboard on the adoption of 3 different APIs (\texttt{Protected Audience}, \texttt{RWS}, and \texttt{Topics}) on a panel of about 60k websites from May 2023 to June 2025~\cite{johnsonAdventPrivacycentricDigital2024,johnsonUnearthingPrivacyEnhancingAd2024}. As far as we can tell, since Sincera's acquisition by another ad tech provider, the data sharing with the researchers has ceased and the online dashboard has been taken down.

\noindent\textbf{Web Almanac.} The second initiative comes from the Web Almanac's yearly reports on the state of the web. These are based on the analysis by volunteers and experts of a single HA monthly crawl. The 2024 and 2025 cookies chapters~\cite{beuginWebAlmanacCookies2024,beuginWebAlmanacCookies2026} presented some results on \texttt{CHIPS}, \texttt{RWS}, and the attestation process. The 2024 privacy chapter~\cite{vekariaWebAlmanacPrivacy2024} touched on a limited analysis of \texttt{Attribution Reporting}, \texttt{FLoC}, \texttt{Protected Audience}, \texttt{Topics}, and \texttt{UA Client Hints}.

\noindent\textbf{Concurrent Work.} Finally, while preparing this submission we came across a much smaller consent-aware measurement than ours from Grib et al.~\cite{gribRiseFallGoogle2026}. The approach in this concurrent work is complimentary to ours on some aspects as it provides pre- and post-consent measurement, validating for instance our observation that some adopters were calling the Privacy Sandbox APIs without user consent. However, the scale of our measurement is unmatched: when we systematically measured \textit{all} different ways to invoke (JS, HTML attributes, and HTTP headers listed in~\autoref{tab:appendix-apis-table}) and use \textit{all} 13 APIs from the Privacy Sandbox, they only partially instrumented a subset of these APIs as reported by~\autoref{tab:prior-work}. Moreover, this concurrent work is only based on data and trends on the top 10k websites and for the last 6 months of 2025 (when adoption rates had long stabilized), when we have access to the \textit{entire} adoption timeline (7+ years from first detection in January 2020 to April 2026) and on the top \textit{100k} websites. As a result, our measurement study remains the largest of its kind and is the only one able to provide a longitudinal view into the prevalence of each feature during the full lifetime of the Privacy Sandbox initiative.

\section{Conclusion}

Our longitudinal results demonstrate how adoption of the newly proposed Privacy Sandbox APIs and technologies remained very limited and uneven across the years.
The adoption rates of the flagship, but now discontinued, \texttt{Attribution Reporting}, \texttt{Protected Audience}, and \texttt{Topics} APIs never reached more than 23.1\%, 5.7\%, and 14\% of page loads, respectively.
For APIs still supported by Chrome, the situation is similarly uneven; adoption rates for \texttt{FedCM}, \texttt{Fenced Frames}, and \texttt{Storage Access} have been lower than 5.2\%, while partitioned cookies (\texttt{CHIPS}) are present on 34.5\% of page loads, but only account for 6.7\% of third-party cookies (TPC) on average, and \texttt{UA Client Hints} API is being accessed on more than 62\% of page loads.

If anything, Google's global experiment with the Privacy Sandbox has taught us that only a few actors are willing to experiment with new proposals that may or not be deployed, and that even then, they tend to focus on the technologies that would affect their main use cases or services such as (re)targeted advertising or conversion measurement.
We can also expect that potential future attempts from Google at disrupting web tracking and deprecating TPC will be met with even more skepticism from the ecosystem knowing that the Privacy Sandbox plans did not go through.

Ultimately, the continuing (ab)use of \texttt{UA Client Hints} by the ecosystem is emblematic of the \emph{cat-and-mouse game} of trackers circumventing privacy protections adopted by browsers and users, and further showcases the need to properly consider the fingerprinting potential of newly introduced web APIs; as expressed by the W3C Privacy Working Group in their guidance note~\cite{w3cprivacyworkinggroupW3cFingerprintingguidance2025}.
Perhaps, more striking are the limitations and disparities across browser vendors of \emph{``fix it in the browser''} remedies that the Privacy Sandbox illustrates once more; while Google's global experiment resulted in largely opt-in only restrictions being implemented in Chrome, stronger privacy protections were enabled by default in the meantime into other browsers (e.g., Brave, Firefox, and Safari).

\section*{Acknowledgment}
\ifanonymous
Anonymized for review.
\else

The authors thank Kyle Domico, Jonathan Gregory, Kunyang Li, Jean-Charles Noirot Ferrand, and Jack West for their feedback on initial versions of this paper.

\noindent \textbf{Funding acknowledgment:}
This material is based upon work supported by the National Science Foundation under Grant No. CNS-2343611. Any opinions, findings, and conclusions or recommendations expressed in this material are those of the author(s) and do not necessarily reflect the views of the National Science Foundation. This work was supported in part by the Semiconductor Research Corporation (SRC) and DARPA.
The data analysis on BigQuery was also made possible by support provided through Google Cloud Research Credits program with the award GCP-012E72-720588-E16BBE.
\fi

\section*{Ethics Considerations}
\label{appendix:ethics-considerations}

This work is a longitudinal measurement of the adoption of Google's Privacy Sandbox APIs based on existing and publicly available datasets that did not require to crawl websites ourselves. Nevertheless, we analyze next the limited ethical implications of our work, before justifying conducting it.

\noindent \textbf{Data Collection.} Our work uses the following datasets:
(1) \textit{Chrome Telemetry:} aggregated and anonymous statistics on features usage publicly released by Google, i.e., with no identifiable information available or recoverable about Chrome users who have telemetry turned on.
(2) \textit{HTTP Archive:} archival crawl results for millions of URLs and pages visited each month from Chrome on a desktop and (emulated) Android devices. This is a well established non-profit project whose collected data to be queried by anyone, this avoids redundant similar crawls of websites by independent actors and allows to study the evolution of the web. The crawl identifies itself by appending a \texttt{PTST} token to its User-Agent in case web servers would like to block it.
(3) \textit{Open-source repositories:} we use public version control information from the \textit{chromium} and \texttt{RWS} repositories to collect the attestation and \texttt{RWS} lists shipped in Chrome.\\

\noindent \textbf{Stakeholder Analysis.} Our work encompasses six stakeholders: (1) \textit{Online users}: are individuals whose web activity is traditionally tracked and profiled. They decide which web browsers, extensions, and other privacy settings to use.
(2) \textit{Browser vendors}: are the developers of user-agents that browse and mediate access to the web on behalf of users. They need to consider which features, proposals, and standards to support and implement while balancing security, privacy, usability, performance, and compatibility issues. Here, Google Chrome is behind the Privacy Sandbox initiative.
(3) \textit{Website publishers}: are in charge of managing websites, maintaining their functionality, and monetizing their content while new privacy and security changes are proposed to the web. Some Privacy Sandbox APIs were intended for website publishers (i.e., first-party).
(4) \textit{Advertising and tracking platforms}: are services that provide online tracking and advertising infrastructures. They are affected by regulatory changes, new browser policies or modifications, and public scrutiny. Several Privacy Sandbox APIs were aimed at being adopted by such third-parties.
(5) \textit{Regulators and policymakers}: are entities that draft, propose, and enforce data protection regulations and privacy laws. They often rely on measurements to better understand ecosystems dynamics. The UK's CMA was investigating Google's proposals and cookies deprecation on Chrome.
(6) \textit{Researchers}: scientists from academia and industry studying modifications to the web and effects on privacy, advertising, regulations, etc.\\

\noindent \textbf{\textit{Positive Impacts of the Research.}}
(1) \textit{Improved Understanding (Impact on all Stakeholders)}: by performing the largest measurement of its kind of the Privacy Sandbox proposals, we shed light on which APIs were used in practice, how, and by which actors over time. This provides the most accurate existing overview of the events and ecosystem dynamics related to the adoption and deprecation of these proposals.
(2) \textit{Future Research Directions (Impact on Browser Vendors, Regulators, and Researchers)}: by discussing the main research challenges in evaluating these proposals, we identify several directions where browser vendors, regulators, and researchers could work on to improve the web in future efforts.
(3) \textit{Informing Platform and Policy Design (Impact on Browser Vendors and Regulators)}: by synthesizing years of developments, this work informs about the main issues and gaps uncovered in the Privacy Sandbox, while suggesting future policy changes---both at browser and legal-levels.\\

\noindent \textbf{\textit{Negative Impacts of the Research.}}
(1) \textit{Potential for Incorrect Detection (Impact on Adopters)}: to detect JavaScript implementations of the Privacy Sandbox APIs, we used a syntactic matching technique that could lead to some false positives if the API method was present in a comment or in a branch not taken by the execution path. Similarly, detections could have been missed if obfuscation techniques were used. To control for these, we only look for JavaScript implementations on pages for which Chrome is already reporting the use of one of the APIs. This reduces the possibility that adopters have been flagged incorrectly or are missing.
(2) \textit{Misinterpretation of Findings (Impact on Policy)}: Policy actors may selectively use parts of our findings to justify particular technical or regulatory choices, potentially ignoring some nuanced trade-offs. Specifically, we caution against interpreting our findings as definitive evidence of non-compliance from some adopters. Indeed, we are not lawyers, the legality of calling these Privacy Sandbox APIs prior to user consent has not been debated as far as we know, and a more controlled experiment (such as using a consent-aware crawler interacting with consent banners) would likely be required to confirm violations.

\noindent \textbf{Justification for Research.}
Despite the prior risks, several factors already explained above also help mitigate the highlighted potential negative impacts. As such, we believe that publishing this work is ethically justified and necessary. The primary ethical risk lies in not reflecting on the changes brought by Google's Privacy Sandbox global experiment and to miss the opportunity to better understand the dynamics in the ecosystem and what happened. By linking our measurement findings to recommendations, we aim to encourage a reflection on lessons that can be learned with respect to solving novel research challenges, building better privacy enhancing technologies, providing more accountability, and enforcing regulations. Overall, this measurement study contributes to an ethically grounded understanding of the web, its evolution, and challenges faced by privacy proposals.

\bibliographystyle{custom-style}
\bibliography{refs}

\newpage
\appendices

\section{Open Science}
\label{sec:open-science}

We release all our collection scripts, extraction queries, datasets, analysis code, and documentation steps as a public artifact to enable availability and reproducibility of our research results.  Our artifact can be accessed at the following URL: \url{https://github.com/privacysandstorm/ps-api-adoption-deprecation}.

To foster future research and dissemination of our results in a more accessible way and to a broader public, we will also host on \url{https://privacysandstorm.com/} an interactive dashboard with graphs and statistics for visitors to explore further our data.
\section{Abbreviations}

\autoref{tab:abbreviations} lists the abbreviations used in this work.

\begin{table}[!ht]
\caption{List of abbreviations used in this paper.}
\label{tab:abbreviations}
\begin{tabular}{@{}cl@{}}
\toprule
\textbf{Symbol} & \textbf{Meaning} \\ \midrule
\texttt{CHIPS} & \texttt{Cookies Having Independent Partitioned State} \\
CMA & Competition and Markets Authority \\
CrUX & Chrome User Experience Report \\
DSP & Demand-side platform\\
\texttt{FedCM} & \texttt{Federated Credential Management} \\
\texttt{FLoC} & \texttt{Federated Learning of Cohorts} \\
GDPR & General Data Protection Regulation \\
HA & HTTP Archive \\
ID & Identifier \\
JS & JavaScript \\
OT & Origin trial (test of a new feature in Chrome) \\
\texttt{RWS} & \texttt{Related Website Sets} \\
SSP & Supply-side platform\\
TPC & Third-party cookies \\
\texttt{UA} & \texttt{User-Agent}  \\ \bottomrule
\end{tabular}
\end{table}

\section{Supplemental Tables and Figures}
\label{appendix:api-details}
\label{sec:supplemental-figures}

\begin{table*}[!ht]
\centering
\fontsize{6.3}{6}
\caption{Permissions policy directives, JS methods, and HTTP headers (if any) measured in this paper for each proposal. The list is also available in our artifact in \texttt{.JSON} format (see~\refappendix{sec:open-science}).}
\label{tab:appendix-apis-table}
\setlength{\tabcolsep}{2pt}
\renewcommand{\arraystretch}{0.15}
\begin{tabularx}{\textwidth}{p{4cm} Y Y }
\toprule
\textbf{\footnotesize Permissions Policy Directives} & \textbf{\footnotesize JS Methods (and HTML Attributes)} & \textbf{\footnotesize HTTP Headers} \\
\midrule
\begin{itemize}[leftmargin=0pt, noitemsep, label={}]
\item \textbf{\footnotesize Attribution Reporting}
\item \texttt{attribution-reporting}
\item \texttt{conversion-measurement}
\end{itemize}
&
\begin{itemize}[leftmargin=0pt, noitemsep, label={}]
\item \texttt{XMLHttpRequest.setAttributionReporting}
\item \texttt{HTMLAnchorElement.attributionSrc}
\item \texttt{HTMLImageElement.attributionSrc}
\item \texttt{HTMLScriptElement.attributionSrc}
\end{itemize}
&
\begin{itemize}[leftmargin=0pt, noitemsep, label={}]
\item \texttt{Attribution-Reporting-Eligible}
\item \texttt{Attribution-Reporting-Register-Source}
\item \texttt{Attribution-Reporting-Register-Trigger}
\end{itemize}
\\
\addlinespace[2pt]\\
\begin{itemize}[leftmargin=0pt, noitemsep, label={}]
\item \textbf{\footnotesize CHIPS}
\end{itemize}
&
&
\begin{itemize}[leftmargin=0pt, noitemsep, label={}]
\item \texttt{Set-Cookie: Secure; Partitioned;}
\end{itemize}
\\
\addlinespace[2pt]\\
\begin{itemize}[leftmargin=0pt, noitemsep, label={}]
\item \textbf{\footnotesize FLoC}
\item \texttt{interest-cohort}
\end{itemize}
&
\begin{itemize}[leftmargin=0pt, noitemsep, label={}]
\item \texttt{document.interestCohort}
\end{itemize}
&
\\
\addlinespace[2pt]\\
\begin{itemize}[leftmargin=0pt, noitemsep, label={}]
\item \textbf{\footnotesize FedCM}
\item \texttt{identity-credentials-get}
\end{itemize}
&
\begin{itemize}[leftmargin=0pt, noitemsep, label={}]
\item \texttt{navigator.credentials.get}
\item \texttt{IdentityCredential.disconnect}
\item \texttt{IdentityProvider.close}
\item \texttt{IdentityProvider.getUserInfo}
\item \texttt{navigator.login.setStatus}
\end{itemize}
&
\begin{itemize}[leftmargin=0pt, noitemsep, label={}]
\item \texttt{Set-Login}
\end{itemize}
\\
\addlinespace[2pt]\\
\begin{itemize}[leftmargin=0pt, noitemsep, label={}]
\item \textbf{\footnotesize Fenced Frames}
\end{itemize}
&
\begin{itemize}[leftmargin=0pt, noitemsep, label={}]
\item \texttt{document.createElement("fencedframe")}
\item \texttt{window.Fence.getNestedConfigs}
\item \texttt{window.Fence.reportEvent}
\item \texttt{window.FencedFrameConfig.setSharedStorageContext}
\item \texttt{window.Fence.setReportEventDataForAutomaticBeacons}
\end{itemize}
&
\begin{itemize}[leftmargin=0pt, noitemsep, label={}]
\item \texttt{Sec-Fetch-Dest: fencedframe}
\item \texttt{Supports-Loading-Mode: fenced-frame}
\end{itemize}
\\
\addlinespace[2pt]\\
\begin{itemize}[leftmargin=0pt, noitemsep, label={}]
\item \textbf{\footnotesize Private Aggregation}
\item \texttt{private-aggregation}
\end{itemize}
&
\begin{itemize}[leftmargin=0pt, noitemsep, label={}]
\item \texttt{privateAggregation.contributeToHistogram}
\item \texttt{privateAggregation.contributeToHistogramOnEvent}
\item \texttt{privateAggregation.enableDebugMode}
\end{itemize}
&
\\
\addlinespace[2pt]\\
\begin{itemize}[leftmargin=0pt, noitemsep, label={}]
\item \textbf{\footnotesize Private State Tokens}
\item \texttt{private-state-token-issuance}
\item \texttt{private-state-token-redemption}
\item \texttt{trust-token-redemption}
\end{itemize}
&
\begin{itemize}[leftmargin=0pt, noitemsep, label={}]
\item \texttt{document.hasPrivateToken}
\item \texttt{document.hasRedemptionRecord}
\item \texttt{XMLHttpRequest.setPrivateToken}
\end{itemize}
&
\begin{itemize}[leftmargin=0pt, noitemsep, label={}]
\item \texttt{Sec-Private-State-Token}
\item \texttt{Sec-Private-State-Token-Lifetime}
\item \texttt{Sec-Private-State-Token-Crypto-Version}
\item \texttt{Sec-Redemption-Record}
\end{itemize}
\\
\addlinespace[2pt]\\
\begin{itemize}[leftmargin=0pt, noitemsep, label={}]
\item \textbf{\footnotesize Protected Audience}
\item \texttt{join-ad-interest-group}
\item \texttt{run-ad-auction}
\end{itemize}
&
\begin{itemize}[leftmargin=0pt, noitemsep, label={}]
\item \texttt{navigator.joinAdInterestGroup}
\item \texttt{navigator.leaveAdInterestGroup}
\item \texttt{navigator.clearOriginJoinedAdInterestGroups}
\item \texttt{navigator.runAdAuction}
\item \texttt{navigator.adAuctionComponents}
\item \texttt{navigator.createAuctionNonce}
\end{itemize}
&
\begin{itemize}[leftmargin=0pt, noitemsep, label={}]
\item \texttt{Ad-Auction-Allowed}
\item \texttt{Ad-Auction-Only}
\item \texttt{Ad-Auction-Signals}
\item \texttt{Ad-Auction-Additional-Bid}
\item \texttt{X-fledge-bidding-signals-format-version}
\item \texttt{Data-Version}
\item \texttt{Sec-Ad-Auction-Fetch}
\end{itemize}
\\
\addlinespace[2pt]\\
\begin{itemize}[leftmargin=0pt, noitemsep, label={}]
\item \textbf{\footnotesize RWS}
\item \texttt{top-level-storage-access}
\end{itemize}
&
&
\\
\addlinespace[2pt]\\
\begin{itemize}[leftmargin=0pt, noitemsep, label={}]
\item \textbf{\footnotesize Shared Storage}
\item \texttt{shared-storage}
\item \texttt{shared-storage-select-url}
\end{itemize}
&
\begin{itemize}[leftmargin=0pt, noitemsep, label={}]
\item \texttt{window.sharedStorage.selectURL}
\item \texttt{window.sharedStorage.run}
\item \texttt{window.sharedStorage.set}
\item \texttt{window.sharedStorage.append}
\item \texttt{window.sharedStorage.delete}
\item \texttt{window.sharedStorage.clear}
\item \texttt{window.sharedStorage.worklet.addModule}
\item \texttt{window.sharedStorage.createWorklet}
\item \texttt{window.sharedStorage.worklet.run}
\item \texttt{window.sharedStorage.worklet.selectURL}
\item \texttt{window.sharedStorage.batchUpdate}
\end{itemize}
&
\begin{itemize}[leftmargin=0pt, noitemsep, label={}]
\item \texttt{Sec-Shared-Storage-Writable}
\item \texttt{Sec-Shared-Storage-Data-Origin}
\item \texttt{Shared-Storage-Write}
\item \texttt{Shared-Storage-Cross-Origin-Worklet-Allowed}
\end{itemize}
\\
\addlinespace[2pt]\\
\begin{itemize}[leftmargin=0pt, noitemsep, label={}]
\item \textbf{\footnotesize Storage Access}
\item \texttt{storage-access}
\end{itemize}
&
\begin{itemize}[leftmargin=0pt, noitemsep, label={}]
\item \texttt{document.requestStorageAccess}
\item \texttt{document.requestStorageAccessFor}
\item \texttt{document.hasStorageAccess}
\item \texttt{document.hasUnpartitionedCookieAccess}
\end{itemize}
&
\begin{itemize}[leftmargin=0pt, noitemsep, label={}]
\item \texttt{Sec-Fetch-Storage-Access}
\item \texttt{Activate-Storage-Access}
\end{itemize}
\\
\addlinespace[2pt]\\
\begin{itemize}[leftmargin=0pt, noitemsep, label={}]
\item \textbf{\footnotesize Topics}
\item \texttt{browsing-topics}
\end{itemize}
&
\begin{itemize}[leftmargin=0pt, noitemsep, label={}]
\item \texttt{document.browsingTopics}
\end{itemize}
&
\begin{itemize}[leftmargin=0pt, noitemsep, label={}]
\item \texttt{Sec-Browsing-Topics}
\item \texttt{Observe-Browsing-Topics}
\end{itemize}
\\
\addlinespace[2pt]\\
\begin{itemize}[leftmargin=0pt, noitemsep, label={}]
\item \textbf{\footnotesize UA Client Hints}
\item \texttt{ch-ua}
\item \texttt{ch-ua-arch}
\item \texttt{ch-ua-bitness}
\item \texttt{ch-ua-form-factors}
\item \texttt{ch-ua-full-version}
\item \texttt{ch-ua-full-version-list}
\item \texttt{ch-ua-high-entropy-values}
\item \texttt{ch-ua-mobile}
\item \texttt{ch-ua-model}
\item \texttt{ch-ua-platform}
\item \texttt{ch-ua-platform-version}
\item \texttt{ch-ua-wow64}
\end{itemize}
&
\begin{itemize}[leftmargin=0pt, noitemsep, label={}]
\item \texttt{navigator.userAgentData.toJSON}
\item \texttt{navigator.userAgentData.getHighEntropyValues}
\end{itemize}
&
\begin{itemize}[leftmargin=0pt, noitemsep, label={}]
\item \texttt{Accept-CH}
\item \texttt{Critical-CH}
\item \texttt{Delegate-CH}
\item \texttt{Sec-CH-UA*}
\end{itemize}
\\
\addlinespace[2pt]\\
\bottomrule
\end{tabularx}
\end{table*}

\begin{figure*}[!ht]
\centering
  \includegraphics[width=.7\linewidth]{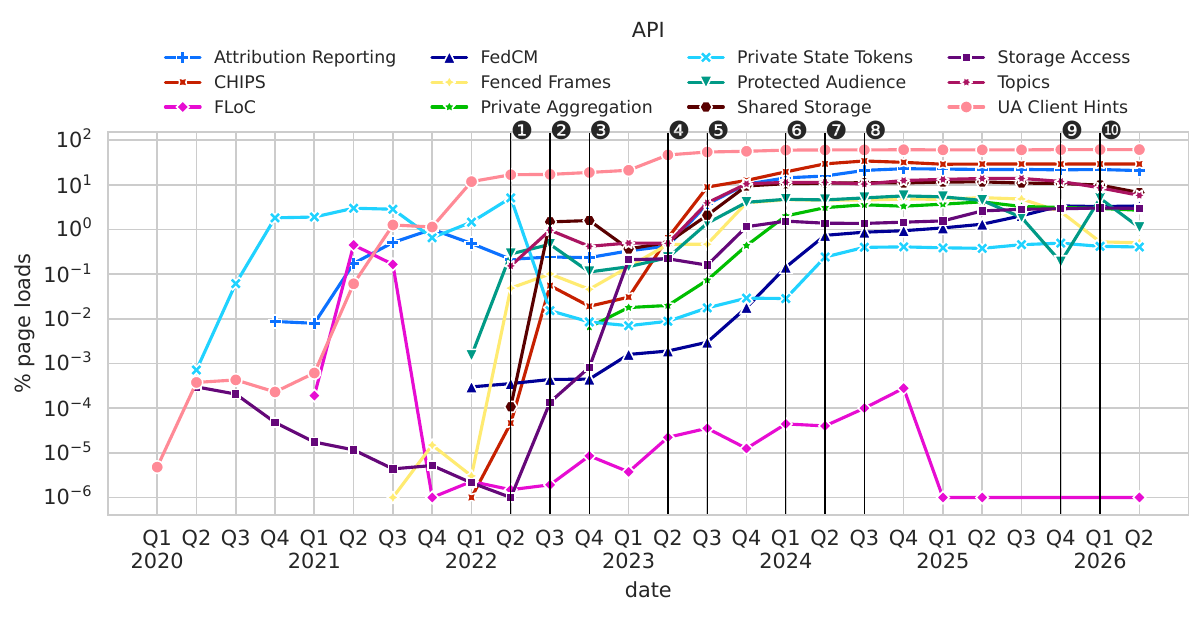}
  \caption{General adoption rates (quarterly percentage of page loads) over time from \textit{Chrome Telemetry} dataset (log-scale version of~\autoref{fig:telemetry}). Vertical numbered lines correspond to major development events (see~\autoref{aside:timeline}).}
  \label{fig:telemetry-log}
\end{figure*}

\begin{figure*}[!ht]
\centering
  \includegraphics[width=.7\linewidth]{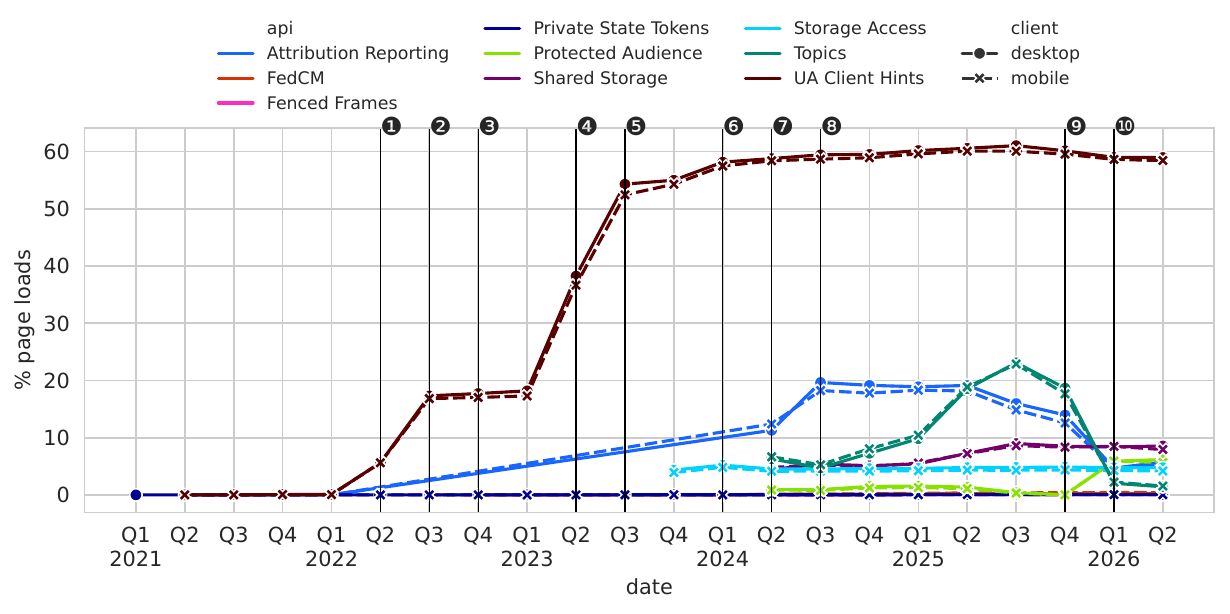}
  \caption{General adoption rates (quarterly percentage of pages) over time from \textit{HA-blink} dataset for desktop and mobile clients (log-scale in~\autoref{fig:ha-blink-log}). Vertical numbered lines correspond to major development events (see~\autoref{aside:timeline}).}
  \label{fig:ha-blink}
\end{figure*}

\begin{figure*}[!ht]
\centering
  \includegraphics[width=.7\linewidth]{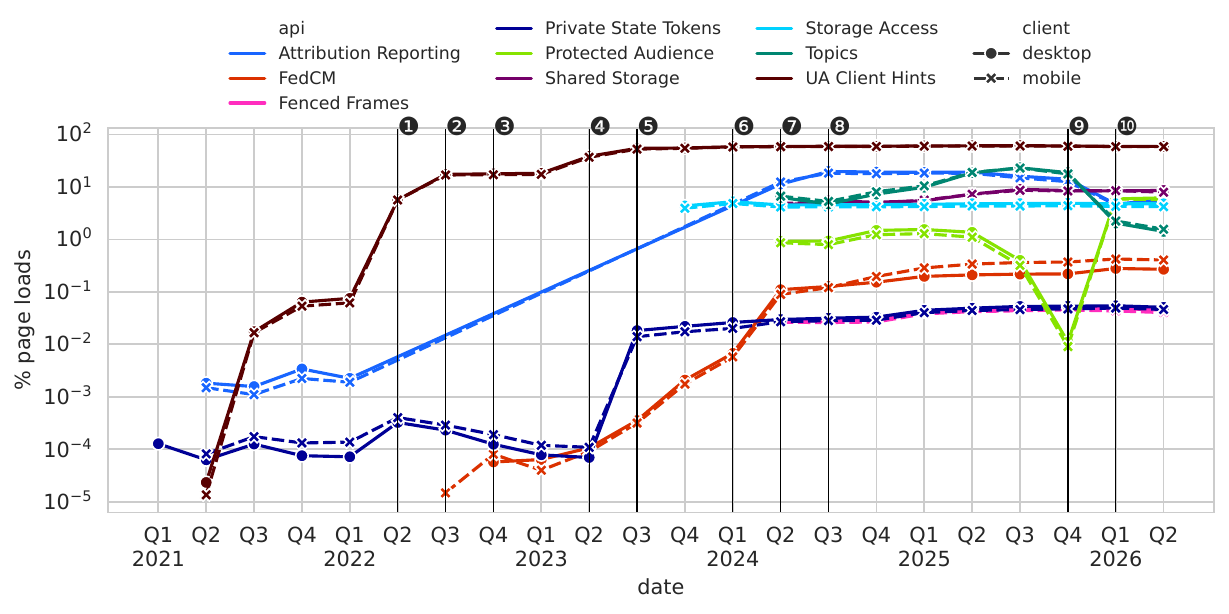}
  \caption{General adoption rates (quarterly percentage of pages) over time from \textit{HA-blink} dataset for desktop and mobile clients (log-scale in~\autoref{fig:ha-blink}). Vertical numbered lines correspond to major development events (see~\autoref{aside:timeline}).}
  \label{fig:ha-blink-log}
\end{figure*}

\end{document}